\documentclass[12pt,a4paper]{article}

\usepackage{epsfig}
\usepackage{amsmath,amsfonts,amssymb}
\usepackage{cite}
\usepackage{colortbl}
\usepackage{pifont}

\providecommand{\openone}{\leavevmode\hbox{\small1\kern-3.8pt\normalsize1}}

\newcommand{\fsm}{f_\text{SM}}

\begin{document}

\begin{center}
\begin{Large}
{\bf Dilepton azimuthal correlations in $t \bar t$ production }
\end{Large}

\vspace{0.5cm}
J.~A.~Aguilar--Saavedra \\
\begin{small}
{ Departamento de F\'{\i}sica Te\'orica y del Cosmos, 
Universidad de Granada, \\ E-18071 Granada, Spain} \\ 

\end{small}
\end{center}

\begin{abstract}
The dilepton azimuthal correlation, namely the difference $\phi$ between the azimuthal angles of the positive and negative charged lepton in the laboratory frame, provides a stringent test of the spin correlation in $t \bar t$ production at the Large Hadron Collider. We introduce a parameterisation of the differential cross section $d\sigma / d\phi$ in terms of a Fourier series and show that the third-order expansion provides a sufficiently accurate approximation. This expansion can be considered as a `bridge' between theory and data, making it very simple to cast predictions in the Standard Model (SM) and beyond, and to report measurements, without the need to provide the numbers for the whole binned distribution. We show its application by giving predictions for the coefficients in the presence of (i) an anomalous top chromomagnetic dipole moment; (ii) an anomalous $tbW$ interaction. The methods presented greatly facilitate the study of this angular distribution, which is of special interest given the $3.2(3.7)\sigma$ deviation from the SM next-to-leading order prediction found by the ATLAS Collaboration in Run 2 data.
\end{abstract}

\section{Introduction}

The production of $t \bar t$ pairs at the large hadron collider (LHC) provides a sensitive probe of the properties of the top quark, both in the production and the decay~\cite{Bernreuther:2008ju,Deliot:2010ey,Aguilar-Saavedra:2014kpa}. 
Among many observables investigated by the ATLAS and CMS Collaborations, the correlation between the spins of the top quark and anti-quark is particularly subtle and difficult to measure. It is well known that the Standard Model (SM) predicts a sizeable $t \bar t$ spin correlation~\cite{Barger:1988jj,Mahlon:1995zn,Stelzer:1995gc}. The spins of $t$ and $\bar t$ are not directly measurable but, due to their short lifetime, they can be accessed through the angular distributions of their decay products. For the decay of a top quark $t \to W^+ b$, $W^+ \to \ell^+ \nu / \bar d u$, with $\ell = e,\mu,\tau$, the decay products have the angular distribution
\begin{equation}
\frac{1}{\Gamma} \frac{d\Gamma}{d\!\cos \theta_i} = \frac{1}{2} \left(1+ P \alpha_i \cos \theta_i \right) \,,
\label{ec:tdec}
\end{equation}
with $\theta_i$ the polar angle between the momentum of the decay product $i = \ell^+, \nu, \bar d, u, b, W^+$ in the rest frame of the parent top quark, and some reference axis $\hat s_t$; $P$ is the top polarisation along that axis, and $\alpha_i$ are constants that, because of angular momentum conservation, must satisfy $|\alpha_i| \leq 1$. For the charged leptons the SM prediction is $\alpha_\ell = 1$ at the tree level, with next-to-leading (NLO) corrections at the permille level~\cite{Brandenburg:2002xr}. Therefore, the correlation between the charged lepton distribution and the top polarisation is (nearly) maximal. Other top quark decay products have smaller spin analysing power, e.g. $\alpha_d = 0.96$, $\alpha_u = -0.32$, $\alpha_b = -0.39$ at NLO. The angular distributions for the decay of a top antiquark are as in (\ref{ec:tdec}) with $\alpha_{\bar i} = \alpha_i$ but reversing the sign of the $\cos \theta_i$ term.

For the production and subsequent decay of a $t \bar t$ pair, the normalised doubly differential cross section reads
\begin{equation}
\frac{1}{\sigma} \frac{d\sigma}{d\!\cos \theta_i \, d\!\cos\theta_j} = \frac{1}{4} \left( 1 - C \alpha_i  \alpha_j \cos \theta_i \cos \theta_j \right) \,,
\label{ec:C}
\end{equation}
with $\theta_i$, $\theta_j$ the polar angles between the momenta of the decay products $i,j$, in the rest frame of the parent top (anti-)quark, and some reference axes $\hat s_t$ and $\hat s_{\bar t}$, respectively. In the above equation we have neglected the small polarisation of $t$ and $\bar t$, which yields terms linear in $\cos \theta_i$ and $\cos \theta_j$. The constant $C$, with $|C| \leq 1$, gives the spin correlation between the top quark and anti-quark for the axes $\hat s_t$ and $\hat s_{\bar t}$.  By choosing orthonormal reference systems in the $t$ and $\bar t$ rest frames, it can be seen that there are nine independent spin correlation coefficients~\cite{Bernreuther:2015yna}, corresponding to various combinations of axes for $t$ and $\bar t$. For example, in the so-called `helicity basis', that is, taking $\hat s_t $ and $\hat s_{\bar t} $ in the direction of the respective top (anti-)quark momenta $\vec k_t$, $\vec k_{\bar t}$ in the $t \bar t$ centre-of-mass (CM) frame, the SM prediction 
at NLO in QCD and electroweak interactions is~\cite{Bernreuther:2013aga,Bernreuther:2015yna} $C_{kk} = 0.310$ at a CM energy of 7 TeV, $C_{kk} = 0.318$ at 8 TeV, and $C_{kk} = 0.331$ at 13 TeV. The nine spin correlation coefficients have been measured by the ATLAS Collaboration at 8 TeV~\cite{Aaboud:2016bit}. Previously, the ATLAS and CMS Collaborations measured $C_{kk}$ in the helicity basis at 7 and 8 TeV~\cite{Chatrchyan:2013wua,Aad:2015bfa,Khachatryan:2016xws}. The measurements are consistent with the SM predictions, see Table~\ref{tab:C}, though the uncertainties are large. 
These uncertainties partly arise from the need to reconstruct the $t$ and $\bar t$ rest frames, as well as the $t \bar t$ CM frame, from their decay products. In the dilepton channel $t \bar t \to \ell^+ \nu b \, \ell^- \bar \nu \bar b$, the reconstruction faces a combinatoric ambiguity due to the two missing neutrinos. In the semileptonic mode $t \bar t \to q \bar q' b \, \ell^- \bar \nu \bar b$, $q=u,c$, $q' = d,s$ (and the charge conjugate decay) with one neutrino the reconstruction is easier but the discrimination between light quarks, based on tracking variables and jet transverse momentum $p_T$, is quite difficult.

\begin{table}[t]
\begin{center}
\begin{tabular}{lcc}
& 7 TeV & 8 TeV \\
ATLAS & $0.315 \pm 0.078$~\cite{Aad:2015bfa} & $0.296 \pm 0.093$~\cite{Aaboud:2016bit} \\
CMS    & $0.08 \pm 0.14$~\cite{Chatrchyan:2013wua} &  $0.276 \pm 0.082$~\cite{Khachatryan:2016xws}
\end{tabular}
\caption{Selected measurements of the spin correlation coefficient $C_{kk}$ in the helicity basis by the ATLAS and CMS Collaborations.}
\label{tab:C}
\end{center}
\end{table}

\begin{figure}[t]
\begin{center}
\includegraphics[height=7cm,clip=]{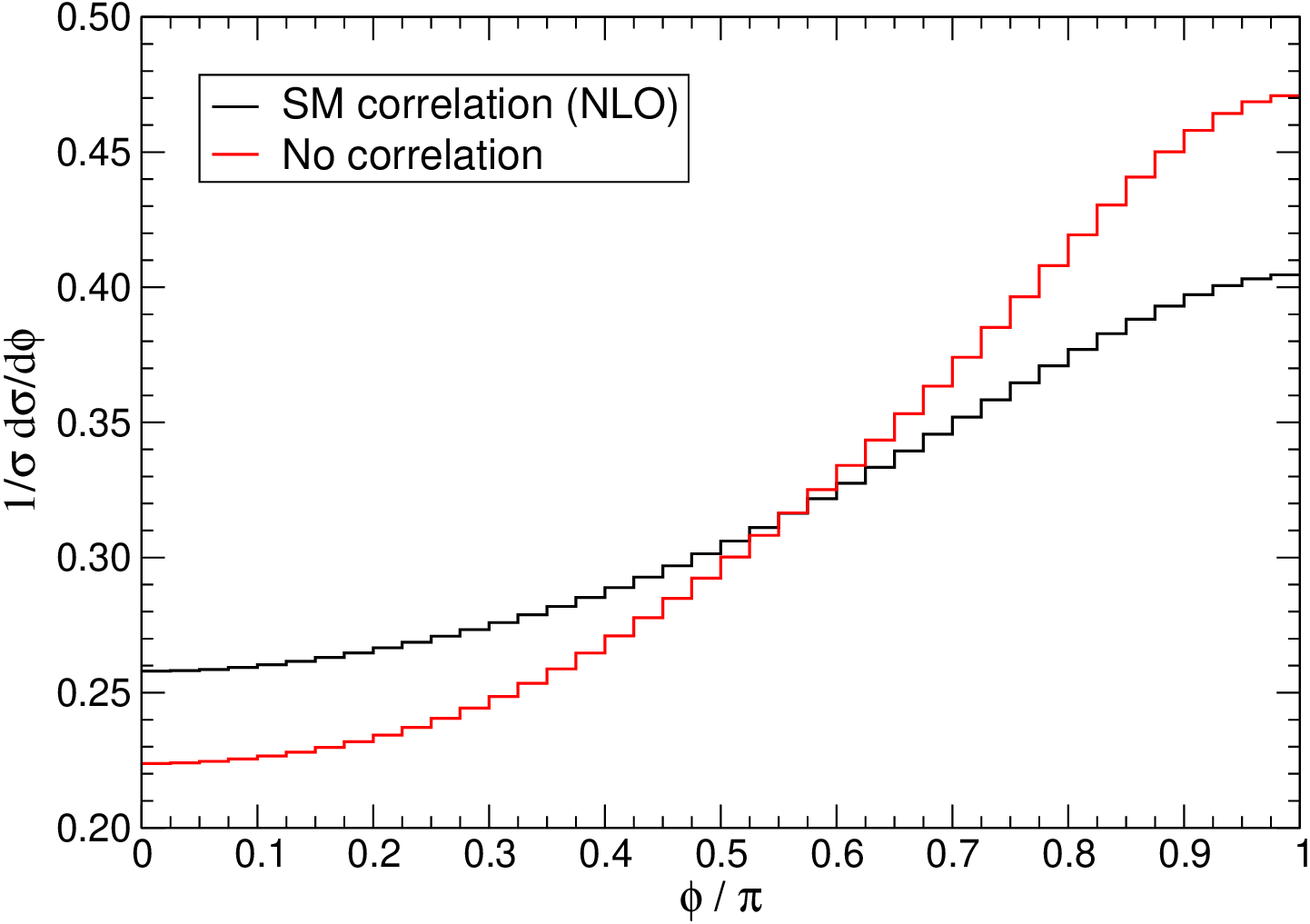}  
\caption{Normalised $d\sigma / d\phi$ distribution (in bins of $\pi/40$) for the SM and the hypothetical case without spin correlation, for a CM energy of 8 TeV.}
 \label{fig:phi8}
\end{center}
\end{figure}

A simpler probe of the $t \bar t$ spin correlation in the dilepton decay mode was pointed out in Ref.~\cite{Mahlon:2010gw}: the laboratory frame dilepton azimuthal correlation, namely the difference $ \phi = | \phi_{\ell^+} - \phi_{\ell^-} |$ between the azimuthal angles of the two charged leptons, taking the $\hat z$ axis in the beam direction. (A predecessor of this correlation was proposed for $Z \to \tau^+ \tau^-$ at the Large Electron Positron Collider, using decay products of the $\tau$ leptons~\cite{Bernabeu:1990na,Alemany:1991ki}.) The $d\sigma / d\phi$ distribution inherits the top spin correlation, and is presented in Fig.~\ref{fig:phi8}, calculated at NLO in QCD interactions for a CM energy of 8 TeV (see the next section for details).
For comparison, we also show the distribution in the absence of spin correlations. This angular distribution allowed to establish the existence of $t \bar t$ spin correlations at the $5.1\sigma$ level already with 7 TeV~\cite{ATLAS:2012ao}. In order to do so, the experimental distribution was compared to a linear combination of the SM and unpolarised one, i.e. the 7 TeV analogues of the distributions shown in Fig.~\ref{fig:phi8}, depending on a parameter $\fsm$,
\begin{equation}
g(\phi;\fsm) \equiv \fsm \left( \frac{1}{\sigma} \frac{d\sigma}{d\phi} \right)_\text{SM} + (1-\fsm)  \left( \frac{1}{\sigma} \frac{d\sigma}{d\phi} \right)_\text{no corr} \,.
\label{ec:fSM}
\end{equation}
The best-fit value of the parameter $\fsm$ was obtained with a likelihood method. The result $\fsm = 1.30 \pm 0.14\;(\text{stat})^{+ 0.27}_{-0.22} \;(\text{sys})$ allowed to exclude the no correlation hypothesis at the $5.1\sigma$ level. Later measurements by the ATLAS and CMS Collaborations have been performed at 7, 8 and 13 TeV~\cite{Aad:2014pwa,Aad:2014mfk,Khachatryan:2016xws,ATLAS:2018rgl} following the same procedure, and the results for $\fsm$ are collected in Table~\ref{tab:fSM}. In particular, the 13 TeV measurement by the ATLAS Collaboration departs $3.2\sigma$ from the NLO prediction ($3.7\sigma$ without including theory uncertainties), following a trend that was already present in earlier measurements but is more apparent in this latest one. 
For the semileptonic $t \bar t$ decay mode a measurement involving the analogous azimuthal angle difference between the charged lepton and the jets has been performed, yielding $\fsm= 1.12 \pm 0.11\;(\text{stat}) \pm 0.22\;(\text{sys})$~\cite{Aad:2014pwa}.

\begin{table}[t]
\begin{center}
\begin{tabular}{lll}
& ATLAS & CMS \\
7 TeV   & $1.19 \pm 0.09\;(\text{stat}) \pm 0.18\;(\text{sys})$~\cite{Aad:2014pwa} & -- \\
8 TeV   & $1.20 \pm 0.05\;(\text{stat}) \pm 0.13\;(\text{sys})$~\cite{Aad:2014mfk}
            & $1.14 \pm 0.06\;(\text{stat})^{+0.15}_{-0.17}\;(\text{sys})$~\cite{Khachatryan:2016xws} \\
13 TeV & $1.250 \pm 0.026\;(\text{stat}) \pm 0.063\;(\text{sys})$~\cite{ATLAS:2018rgl} & -- 
\end{tabular}
\caption{Measurements of the best-fit parameter $\fsm$ in (\ref{ec:fSM}) in the $t \bar t$ dilepton decay mode by the ATLAS and CMS Collaborations.}
\label{tab:fSM}
\end{center}
\end{table}

\begin{figure}[t]
\begin{center}
\includegraphics[height=7cm,clip=]{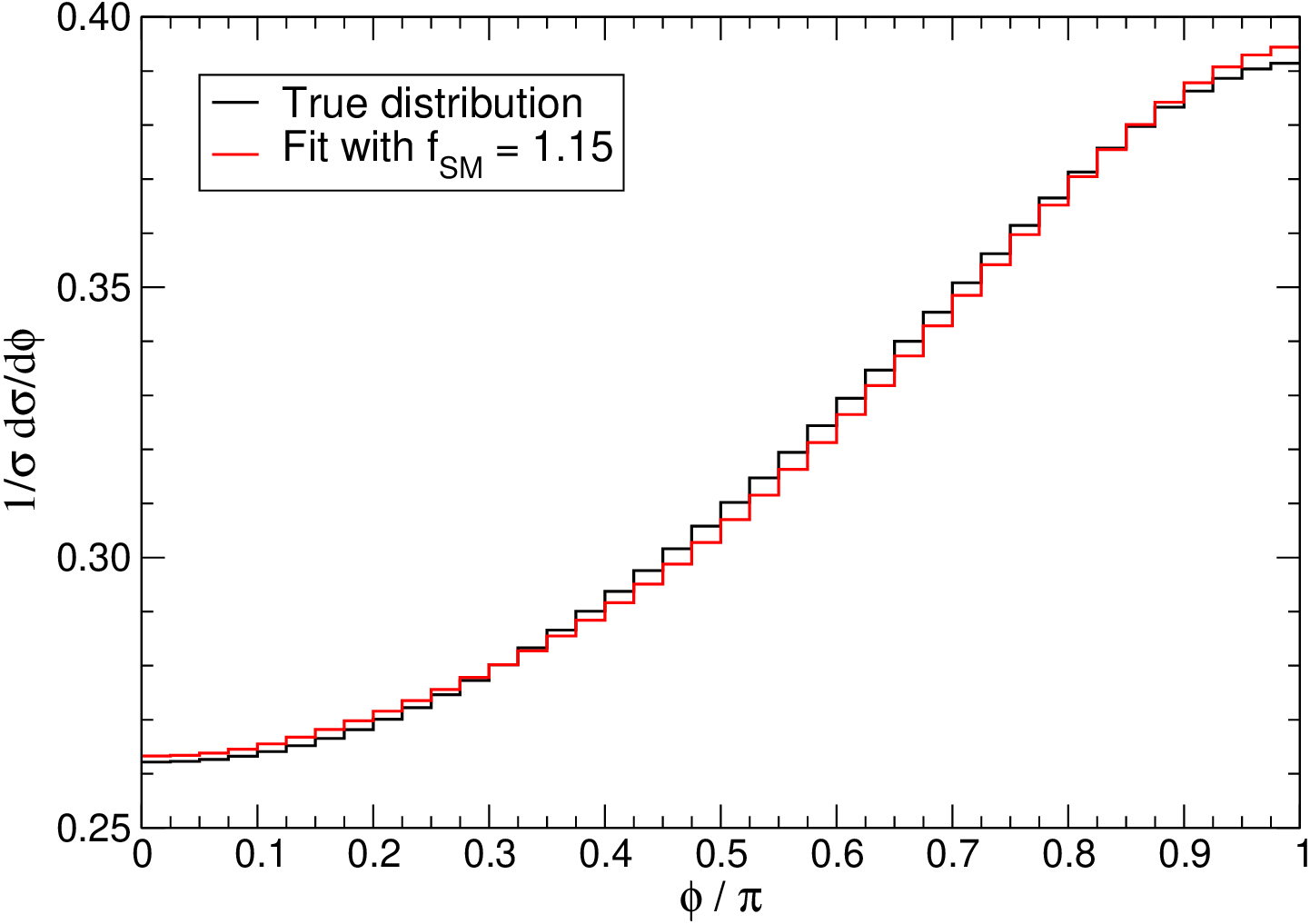}  
\caption{Normalised $d\sigma / d\phi$ distribution (in bins of $\pi/40$)  for a non-SM spin correlation yielding $\fsm = 1.15$ in Eq.~(\ref{ec:fSM}), and best-fit function $g(\phi;\fsm)$ with $\fsm = 1.15$. The CM energy is 8 TeV.}
 \label{fig:distvsfit}
\end{center}
\end{figure}

Using 8 TeV data, the CMS Collaboration has placed limits on new physics directly from the shape of the normalised distribution~\cite{Khachatryan:2016xws}, using the SM prediction at NLO and the first-order contribution from an anomalous top chromomagnetic moment, calculated at leading order (LO)~\cite{Bernreuther:2013aga}. 
The same has been done in Run 2 at 13 TeV, but using also the normalisation as well as the shape~\cite{CMS:2018nzx}.
However, using directly the binned distributions for the theory predictions and to compare with experimental measurements is impractical and difficult to reproduce, for example if one wants to set limits on other types of new physics from experimental data. Instead, it is very convenient to provide the predictions and results in terms of a few numbers,
which can then be compared to test the consistency of the SM with data and set limits on new physics. Clearly, the parameter $\fsm$ in (\ref{ec:fSM}) is not suitable for that, despite its usefulness in establishing the existence of $t \bar t$ spin correlations. First, because the linear combinations $g(\phi;\fsm)$ cannot parameterise {\it any} normalised $d\sigma/d\phi$ distribution, namely, not all possible distributions can be written in the form (\ref{ec:fSM}). In order to make this statement apparent, we generate a  distribution with non-SM spin correlation by injecting a top chromomagnetic dipole moment (see section~\ref{sec:3} below for details) that yields $\fsm = 1.15$, and compare it in Fig.~\ref{fig:distvsfit} with the best-fit function $g(\phi;\fsm)$ with $\fsm = 1.15$. 

A second reason that disfavours $\fsm$ as a parameter to report the measurements is that its experimental determination relies on {\it two} theory predictions, with their corresponding uncertainties: for the SM and for the hypothetical uncorrelated $t \bar t$ production. It is clearly preferrable to provide the result of experiments via theory-independent measurements, and subsequently compare them to the predictions.
We note that, since the nine independent $C$ spin correlation coefficients fully determine the $t \bar t$ spin correlation, the dilepton azimuthal correlation must depend on them.\footnote{Because this distribution in principle depends on all the nine spin correlation coefficients $C$, and not only the one for the the helicity basis, using a measurement of $\fsm$ from the dilepton azimuthal correlation to determine the latter, as it has often been done in the literature, is conceptually incorrect.} However, the dependence may be quite complicated, since it involves boosts from the $t$ and $\bar t$ rest frames to the laboratory frame. It is then more convenient to find a direct, fully general parameterisation of the $d\sigma/ d\phi$ distribution. This will be our task in section~\ref{sec:2}, where we show that a Fourier series expansion up to third order suffices to accurately reproduce the actual distributions. Needless to say, the Fourier coefficients are independent of the particular binning used to report the measured distributions, then they make it very easy to compare results from the ATLAS and CMS experiments, as well as to compare these results with theoretical predictions. 

Ultimately, one is also interested in how the dilepton azimuthal correlation can constrain (or be a signal of) possible new physics effects. This can easily be accomplished with a theoretical calculation of the dependence of the Fourier coefficients on the coupling(s) of the new physics, as we show in section~\ref{sec:3} with the example of anomalous top chromomagnetic moments and in section~\ref{sec:4} with the example of an anomalous $tbW$ interaction. With a set of three functions, giving the dependence of the Fourier coefficients on the new physics coupling(s), one can easily reproduce the prediction for the whole distribution including new physics effects. In section~\ref{sec:a} we use this framework to address the 13 TeV deviation in detail, in the context of an anomalous top chromomagnetic moment, addressing the interplay between the azimuthal correlation and other observables like the total cross section and spin correlation coefficients.

A series expansion of an angular distribution is quite useful as a bridge between theory and data --- provided a small subset of coefficients can accurately reproduce the distribution --- but also provides a bonus: it allows to investigate subtle effects in experimental data that might manifest in the higher-order coefficients. This type of tests can be performed not only on the $d\sigma / d\phi$ distribution on which we focus here, but on any angular distribution, not only in order to probe the presence of new physics indirect effects, but also to test the modeling of the signal, the unfolding procedure, etc. In our discussion in section~\ref{sec:5} we briefly comment on this issue.

\section{Deconstructing the azimuthal correlation}
\label{sec:2}

The distribution $d\sigma / d\phi$ with $ \phi = |\phi_{\ell^+} - \phi_{\ell^-}|$ is defined in the interval $[0,\pi]$. One may extend it to $[-\pi,\pi]$ by taking $ \phi = \phi_{\ell^+} - \phi_{\ell^-}$, as some authors do, in which case it would be symmetric around zero in this interval. Therefore, the Fourier expansion of these distributions only contain cosines,
\begin{equation}
 \frac{1}{\sigma} \frac{d\sigma}{d\phi} = a_0 + \sum_{n=1}^\infty a_n \cos n \phi \,.
 \label{ec:fou}
\end{equation}
The constant term is the overall normalisation. In our case, since $\phi \in [0,\pi]$, we have $a_0 = 1/\pi$ for the normalised distribution.

We calculate the coefficients in the expansion (\ref{ec:fou}) in the SM at NLO in QCD using {\scshape MadGraph5}~\cite{Alwall:2014hca} with NNPDF 2.3~\cite{Ball:2012cx} parton density functions (PDFs), setting dynamic factorisation and renormalisation scales equal to the total transverse mass, $Q = \sum_i m_{Ti}$, with the transverse mass defined as $m_T = \sqrt{m^2 + p_T^2}$. The scale uncertainty is estimated by using twice and one half of this value.  The samples generated have $10^6$ events; the number of positive weight events minus the number of negative weight events is around $6.6 \times 10^5$. (This would be the typical size of data samples after event selection in the dilepton channel for 50 fb$^{-1}$ at 13 TeV.) The Monte Carlo statistical uncertainty is estimated by generating two independent samples. 
Results for the first coefficients, at CM energies of 8 TeV and 13 TeV, are collected in Table~\ref{tab:a}. For comparison, we also show the coefficients for the SM at LO (using samples of $5 \times 10^5$ events) and hypothetical unpolarised case. The latter are calculated using {\scshape MCFM}~\cite{Campbell:2010ff} and the uncertainty quoted is from Monte Carlo statistics only.

\begin{figure}[t]
\begin{center}
\includegraphics[height=7cm,clip=]{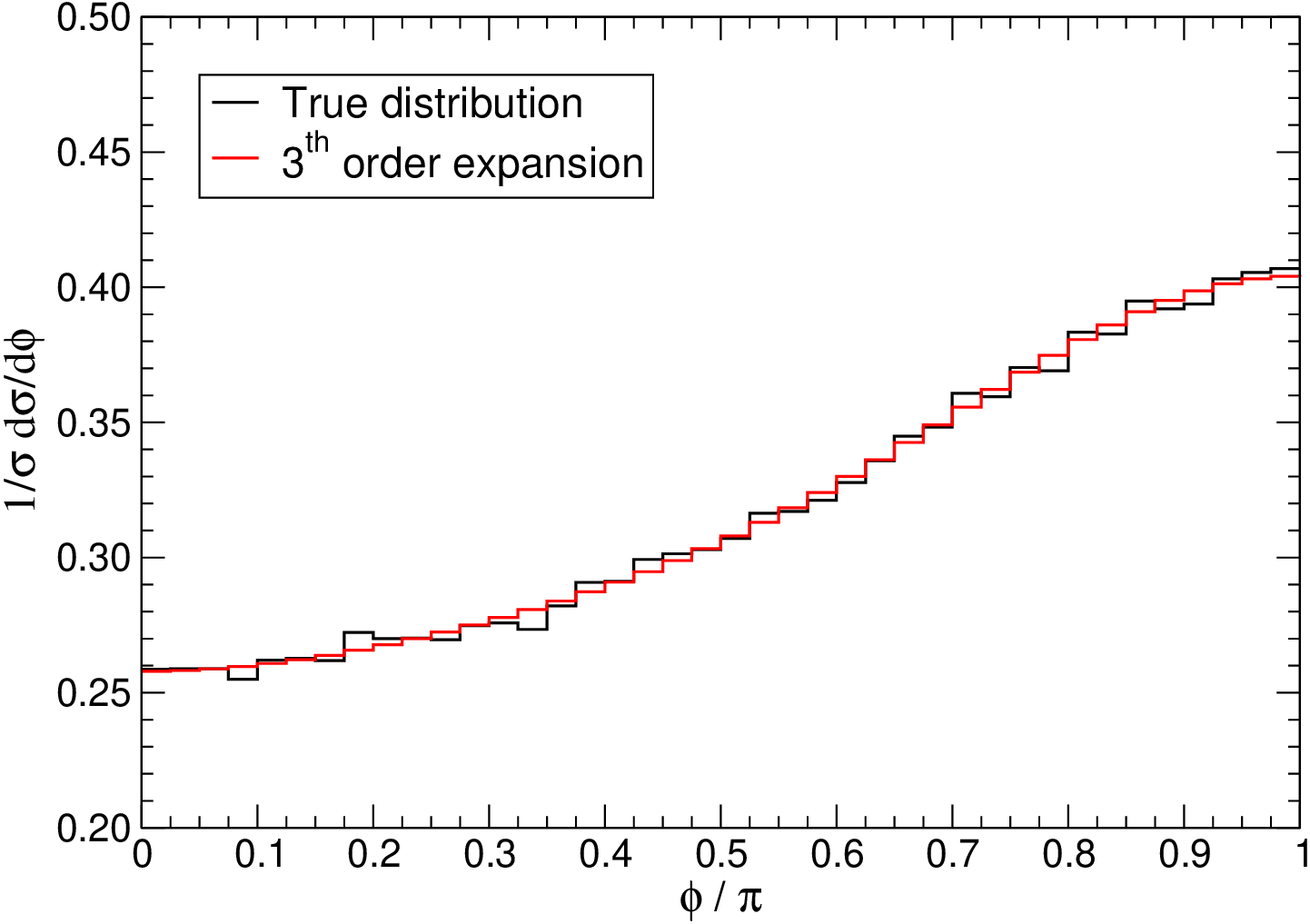}  
\caption{Normalised $d\sigma / d\phi$ distribution (in bins of $\pi/40$)  obtained from Monte Carlo at 8 TeV (black), compared to the third order expansion with the coefficients in Table~\ref{tab:a}.}
 \label{fig:dist3term}
\end{center}
\end{figure}

\begin{table}[htb]
\begin{center}
\begin{tabular}{lccc}
8 TeV & NLO & LO & Uncorrelated $t \bar t$ \\
$a_1$ & $-0.0699^{+0.0014}_{-0.0011}$ 
           & $-0.0762^{+0.0016}_{-0.0022}$
           & $-0.1156 \pm 0.0006$  \\ 
$a_2$ & $0.0127^{+0.0003}_{-0.0002}$  
           & $0.0121^{+0.0026}_{-0.0002}$
           & $ 0.0256 \pm 0.0003$   \\
$a_3$ & $(-3.3 \pm 0.3) \times 10^{-3}$ 
           & $(-4.0 \pm 0.5) \times 10^{-3}$ 
           & $-0.0071 \pm 0.0007$    \\
$a_4$ & $(5.3 \pm 8.4) \times 10^{-4}$ 
           & $(1.6 \pm 0.8) \times 10^{-3}$ 
           & $0.0035 \pm 0.0014$ \\[5mm]
\hline \\
13 TeV & NLO & LO & Uncorrelated $t \bar t$ \\
$a_1$ & $-0.0764^{+0.0023}_{-0.0012}$
           & $-0.0842^{+0.0004}_{-0.0009}$
           & $-0.1187\pm 0.0002 $ \\ 
$a_2$ & $0.0151^{+0.0006}_{-0.0004}$
           & $0.0172 \pm 0.0004$
           & $ 0.0275\pm 0.0003 $ \\
$a_3$ & $(-4.0 \pm 1.1) \times 10^{-3}$
           & $(-4.4 \pm 1.1) \times 10^{-3}$ 
           & $-0.0075 \pm 0.0002 $  \\
$a_4$ & $(1.78 \pm 1.0) \times 10^{-3}$
           & $(1.6 \pm 1.0) \times 10^{-3}$ 
           & $0.0022 \pm 0.0004 $
\end{tabular}
\caption{Lowest order coefficients in the expansion (\ref{ec:fou}) of the normalised $d\sigma / d\phi$ distribution, for CM energies of 8 TeV (up) and 13 TeV (down).}
\label{tab:a}
\end{center}
\end{table}
The `effective' spin correlation, that is, the slope of the distribution (approximately represented by the best-fit parameter $\fsm$) mainly depends on the first non-trivial coefficient $a_1$. The effect of $a_2$ is small, and the influence of $a_3$ and $a_4$ is marginal. This also happens when including new physics contributions of a moderate size in the production or the decay.
For $a_4$ the uncertainty given in Table~\ref{tab:a} is dominated by the Monte Carlo statistics. At both energies this coefficient and higher-order ones are very small, therefore the distributions are well approximated by the third-order expansion, and we will do so in the following.
In Fig.~\ref{fig:dist3term} we compare the actual distribution obtained from the Monte Carlo simulation for the SM at 8 TeV, with the third-order expansion with the coefficients in Table~\ref{tab:a}, finding very good agreement.

We have explored other possibilities for the parameterisation of the distributions. One obvious candidate would be an expansion in terms of Legendre polynomials,
\begin{equation}
\frac{1}{ \sigma} \frac{d\sigma}{dx} = \sum_{l=0}^\infty b_l P_l (x) \,,\quad x = \cos \phi \,.
\label{ec:leg}
\end{equation}
This type of expansion was already used by the CDF Collaboration to investigate the anomalous forward-backward asymmetry observed in $t \bar t$ production at the Tevatron~\cite{CDF:2013gna}.
However, because $d\sigma /dx \sim 1/\sqrt{1-x^2}$, the distribution does not admit a series expansion of this type. (The function $f(x) = 1/\sqrt{1-x^2}$ does not belong to the Hilbert space $L^2([-1,1])$ of quadratically integrable functions.)  
One can get rid of this difficulty by modifying the expansion as
\begin{equation}
\frac{1}{ \sigma} \frac{d\sigma}{d\phi} = \sum_{l=0}^\infty b_l P_l (x) \,.
\label{ec:leg2}
\end{equation}
This is equivalent to the Fourier series (\ref{ec:fou}) we have used, as it can easily be shown using trigonometric identities, but more complicated because the normalisation of the distribution is not only determined by the first coefficient $b_0$, but by a combination of the even coefficients, $1 = \pi \, b_0 + \pi/4 \, b_2 + 9 \pi/64 \, b_4 + \cdots$. Therefore, the simpler expansion (\ref{ec:fou}) is preferred.

\section{Effect of a top chromomagnetic moment}
\label{sec:3}

As an example of new physics in $t \bar t$ production that modifies the spin correlation we consider a top chromomagnetic moment. The $ttg$ interaction, including the SM as well as the contribution from gauge-invariant dimension-six operators, can be written as~\cite{AguilarSaavedra:2008zc}
\begin{equation}
  \mathcal{L}_{ttg} \!=\!
    - g_s \bar t \gamma^\mu \frac{\lambda^a}{2} t \, G_\mu^a
    - \frac{g_s}{m_t} \bar t \sigma^{\mu \nu} (d_V + i d_A \gamma_5)
         \frac{\lambda^a}{2} t\,  G_{\mu \nu}^a \,
\label{ec:lagr}
\end{equation}
in standard notation, with $d_V$ and $d_A$ the top chromomagnetic and chromoelectic dipole moments respectively, $G_{\mu \nu}^a$ the gluon field strength tensor, $\lambda^a$ the Gell-Mann matrices, $g_s$ the strong coupling constant and $m_t$ the top quark mass. The second term contains both $t t g$ and $ttgg$ interactions and can arise from the dimension-six operator~\cite{Buchmuller:1985jz}
\begin{equation}
 O_{uG\phi}^{33} = (\bar q_{L3} \lambda^a \sigma^{\mu \nu} t_R) \tilde \phi
  \, G_{\mu \nu}^a\ ,
\end{equation}
with $q_{L3} = (t_L \, b_L)^T$, $\phi$ the Higgs doublet and $\tilde \phi = i \tau_2 \phi^*$. Anomalous moments $d_V$, $d_A$ can be constrained from the measurements of inclusive cross sections~\cite{Haberl:1995ek,Hioki:2009hm,Hioki:2013hva,Barducci:2017ddn}, differential distributions~\cite{Cheung:1995nt,HIOKI:2011xx,Kamenik:2011dk,Aguilar-Saavedra:2014iga,Franzosi:2015osa}, and the $t \bar t$ spin correlation~\cite{Bernreuther:2013aga}. For simplicity we will set $d_A = 0$ and study the influence of a non-zero $d_V$ on the dilepton azimuthal correlation. In the SM a chromomagnetic moment $d_V = 0.007$ is generated at the one loop level, mainly arising from QCD corrections to the $ttg$ vertex~\cite{Martinez:2007qf}. We ignore it in our calculations, as these QCD corrections are already included in our NLO calculation for $pp \to t \bar t$, and only consider anomalous contributions to the second term in (\ref{ec:lagr}).

\begin{figure}[htb]
\begin{center}
\begin{tabular}{ccc}
\includegraphics[height=4.8cm,clip=]{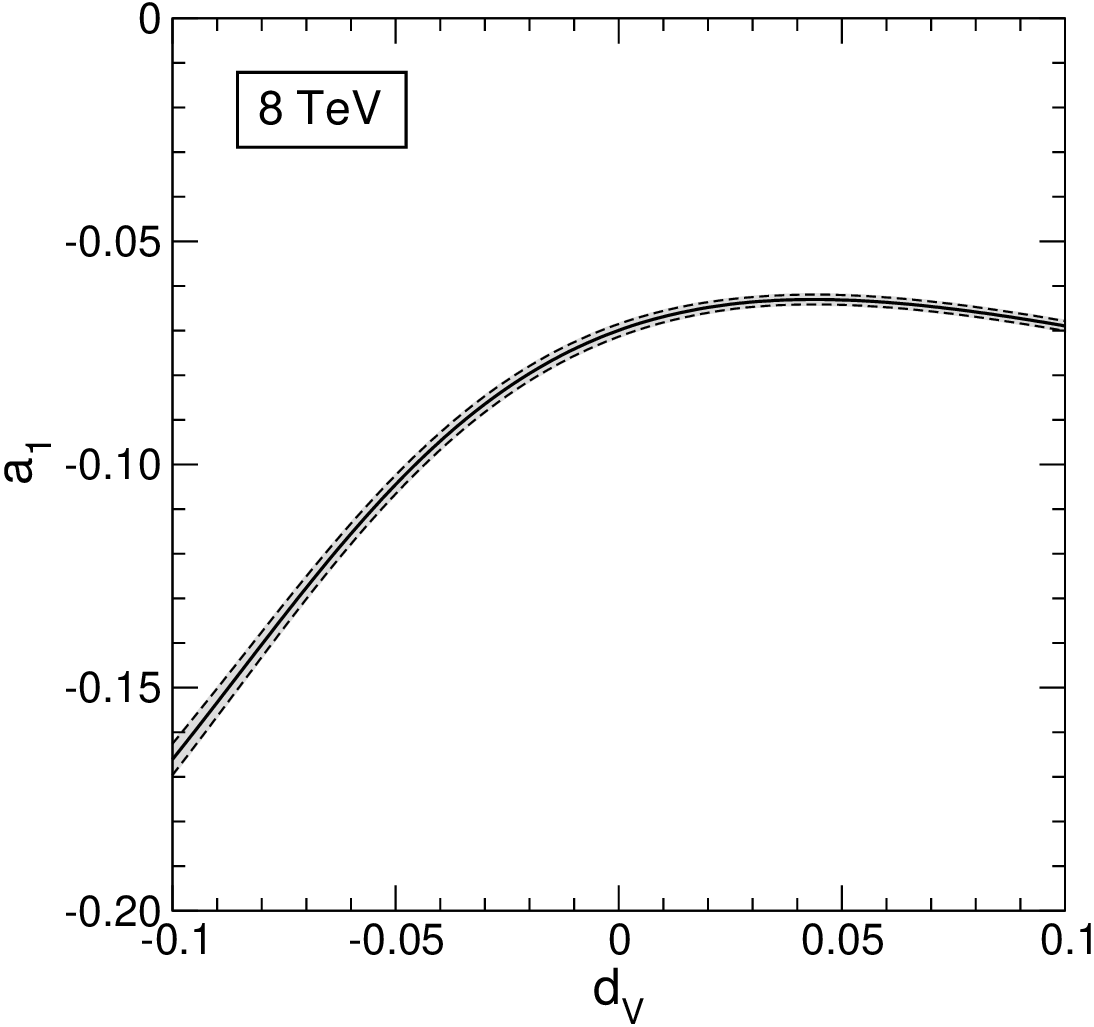}  &
\includegraphics[height=4.8cm,clip=]{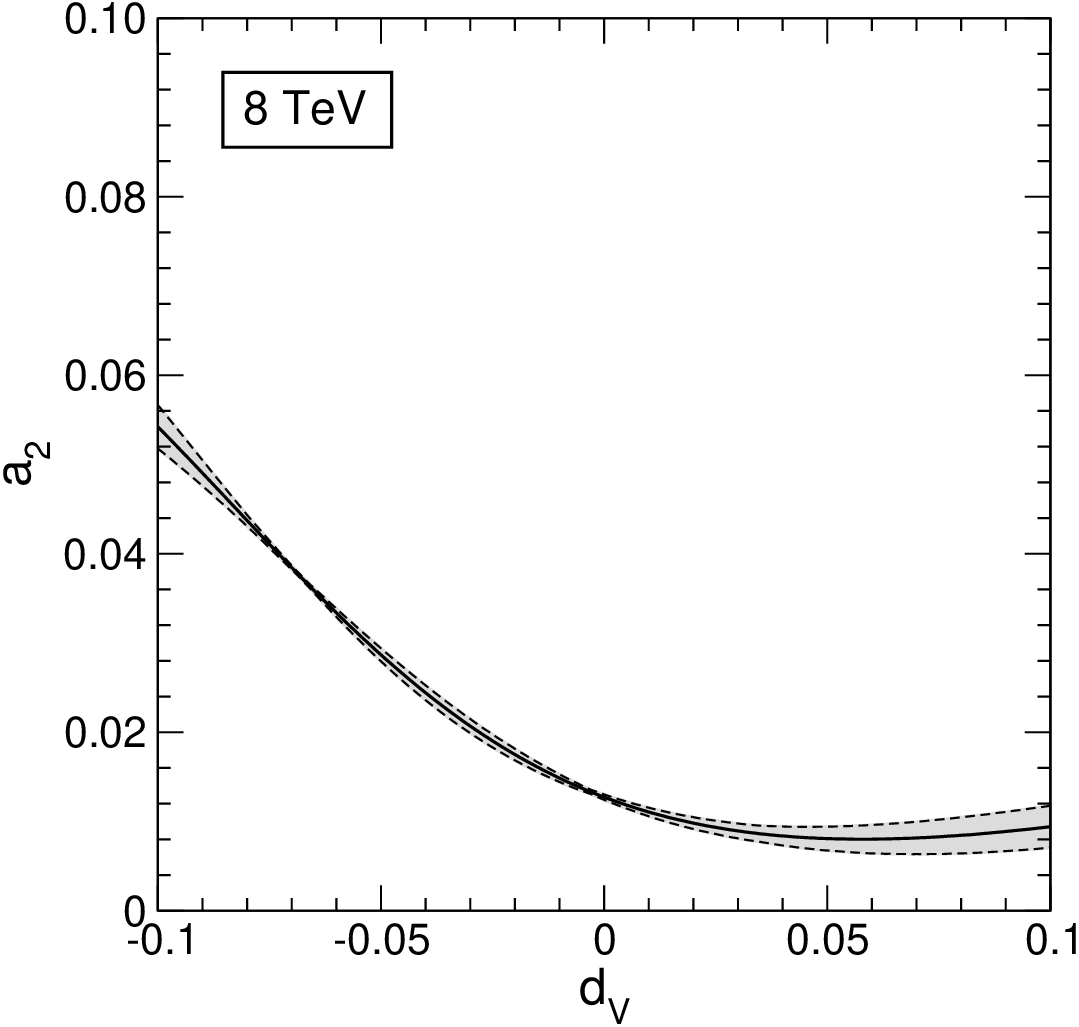}  &
\includegraphics[height=4.8cm,clip=]{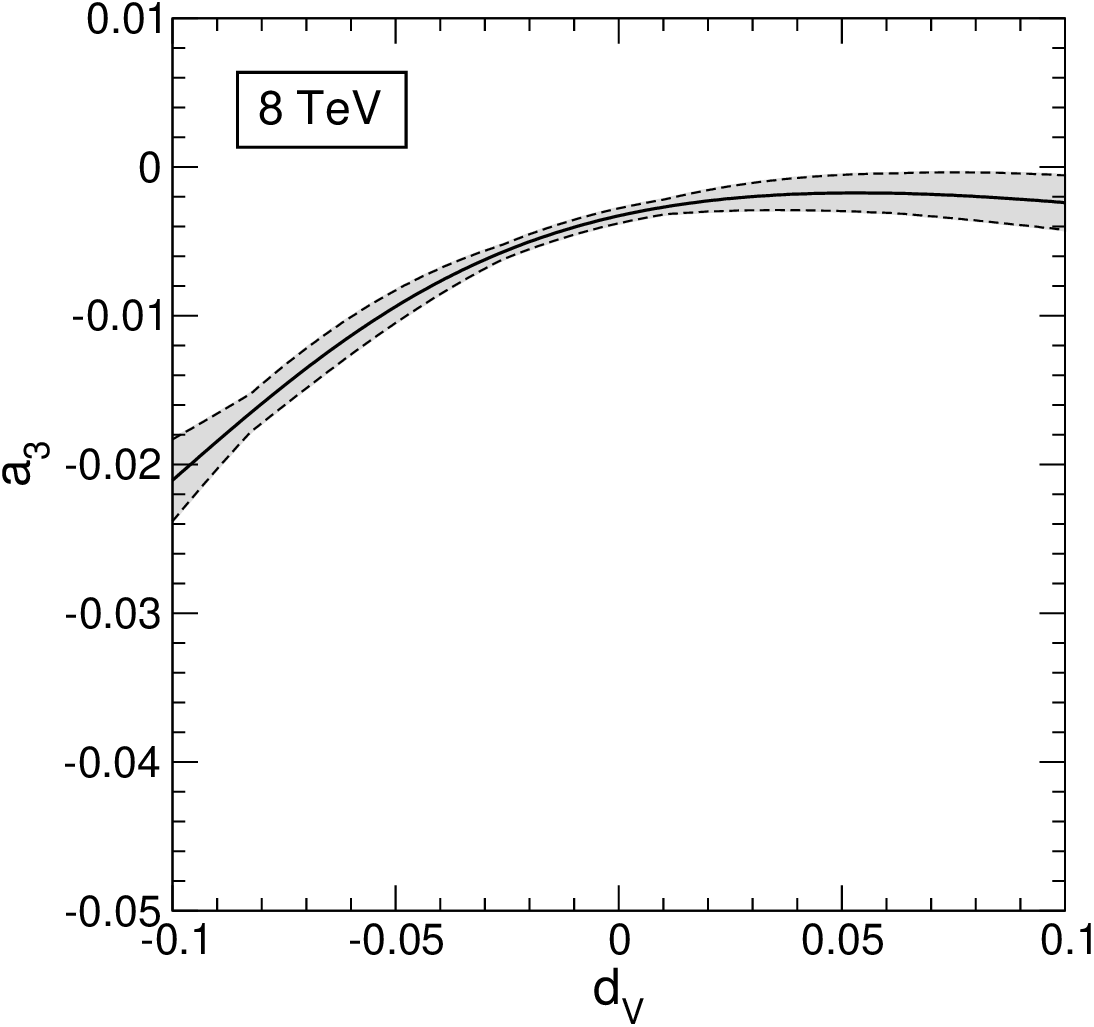}  \\
\includegraphics[height=4.8cm,clip=]{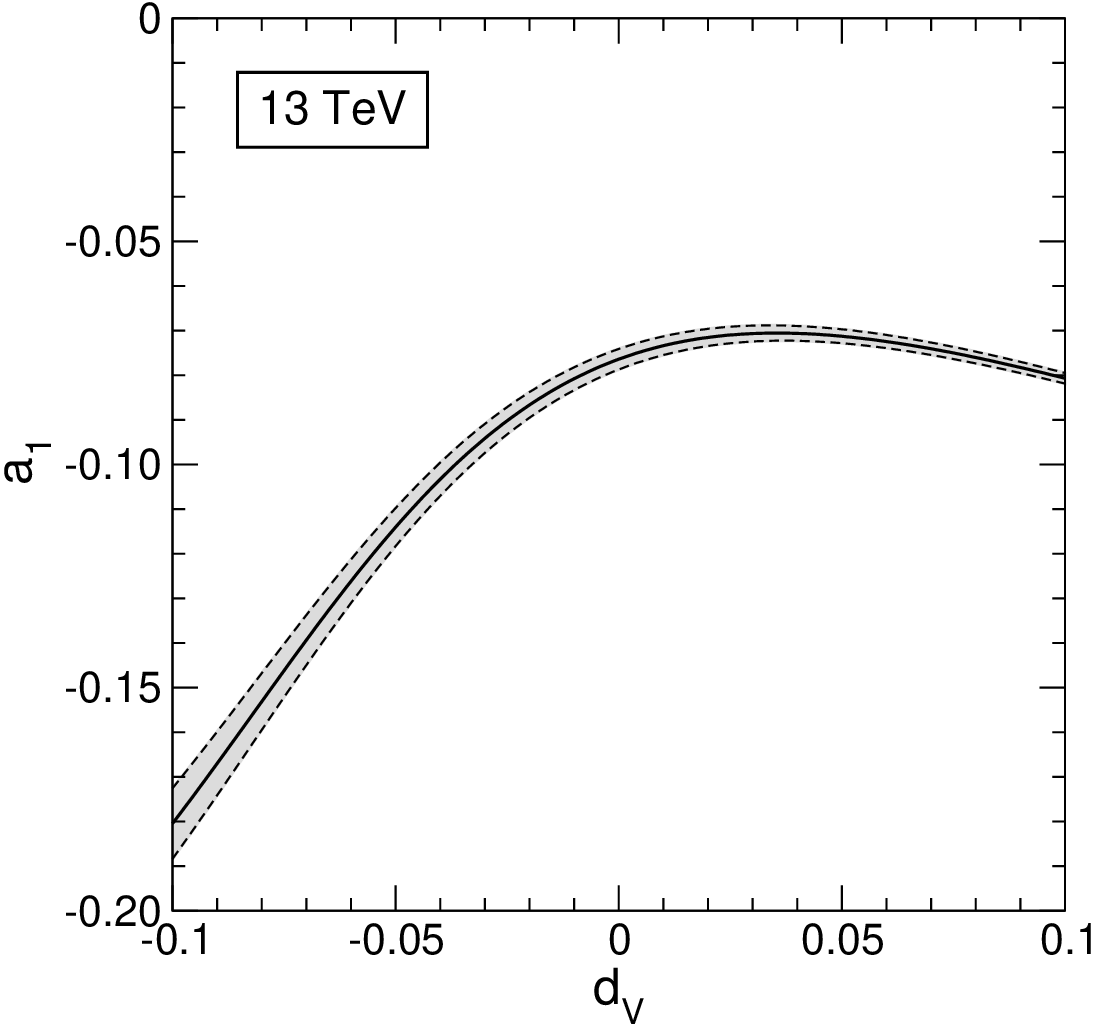}  &
\includegraphics[height=4.8cm,clip=]{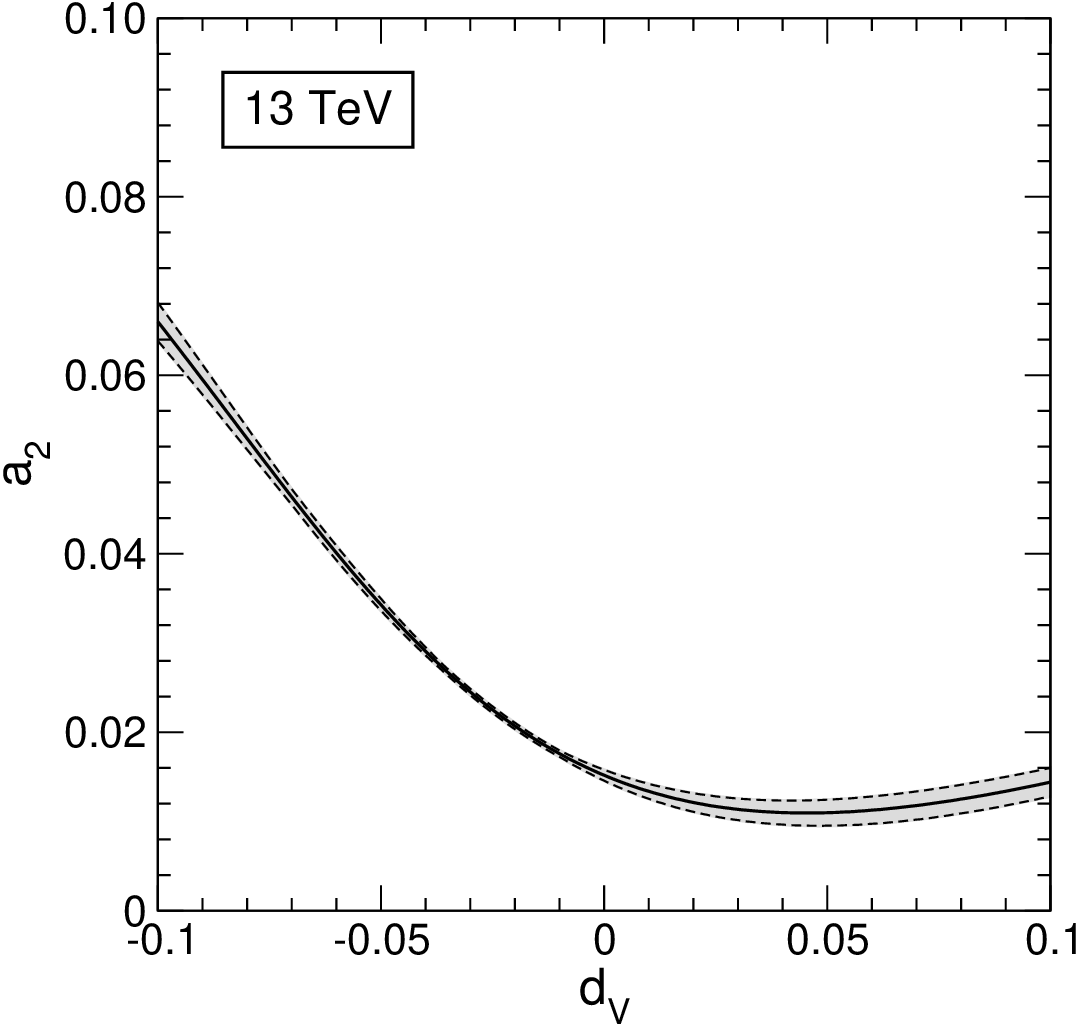}  &
\includegraphics[height=4.8cm,clip=]{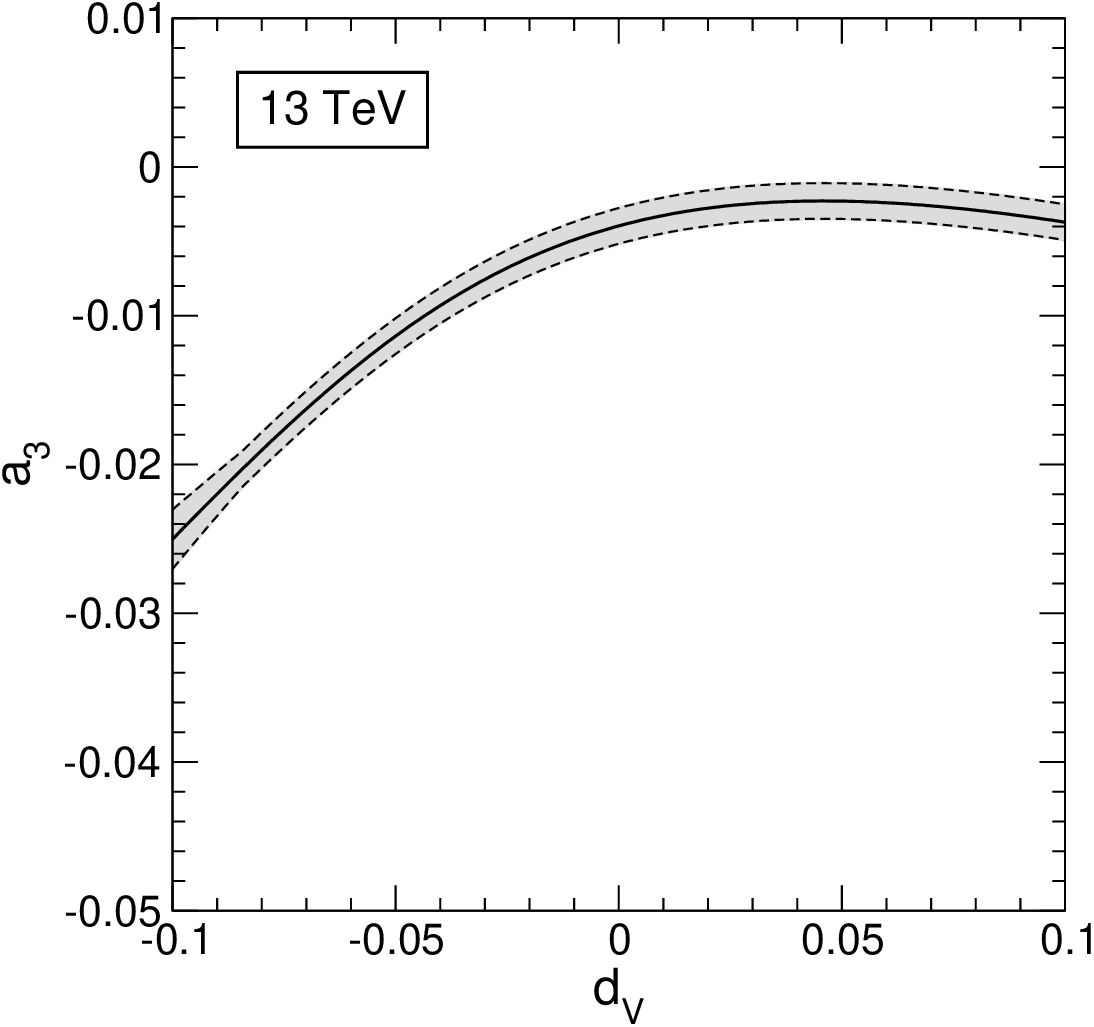}  
\end{tabular}
\caption{Dependence of the first Fourier coefficients in (\ref{ec:fou}) on a possible top chromomagnetic moment $d_V$, for CM energies of 8 TeV and 13 TeV.}
 \label{fig:avsdV}
\end{center}
\end{figure}

The dipole interactions enter at most twice the amplitudes for $t \bar t$ production, therefore the cross section depends quadratically on $d_V$. The dependence of the Fourier coefficients in (\ref{ec:fou}) on $d_V$ can be obtained with a simple procedure. One first considers the unnormalised distribution
\begin{equation}
 \frac{d\sigma}{d\phi} = \bar a_0 + \sum_{n=1}^\infty \bar a_n \cos n \phi \,,
 \label{ec:fou2}
\end{equation}
with $\bar a_0 = \sigma/\pi$. Because the functions $\cos n \phi$ are orthogonal in $[0,\pi]$, with $\int_0^\pi \cos^2 n \phi \, d\phi = \pi/2$, the coefficients can be obtained from a sample of $N$ unweighted events as
\begin{equation}
\bar a_0 = \sigma / \pi \,, \quad \bar a_n=  \frac{2 \sigma}{\pi N} \sum_j \cos n \phi_j \,, \forall n > 0 \,,
\end{equation}
with $j$ running over the events and $\phi_j$ the corresponding value of $\phi$.
By generating event samples for different values of $d_V$, and extracting $\bar a_n$ for each sample, their functional dependence $\bar a_n(d_V)$, which is a fourth-order polynomial too, can be determined. The coefficients of the normalised distribution are
\begin{equation}
a_n(d_V) = \frac{\bar a_n(d_V)}{ \pi \bar a_0(d_V)} \,.
\end{equation}
Our calculations are performed including the SM NLO contribution, the interference between the LO SM and new physics, and the pure new physics contributions at LO. This is the approach taken in Ref.~\cite{Bernreuther:2013aga}, with the difference that we use non-expanded denominators when computing the normalised $a_n$, and keep terms beyond the linear order in $d_V$. There are several arguments~\cite{Aguilar-Saavedra:2014kpa,Contino:2016jqw,AguilarSaavedra:2018nen} that justify keeping all the terms even if dimension-eight operators are not included.
 At variance with Ref.~\cite{Bernreuther:2013aga}, we also include a  factor $K = \sigma_\text{SM}^\text{NLO} / \sigma_\text{SM}^\text{LO}$ in the LO calculations of the new physics contribution and its interference with the SM, in order to improve the approximation and mimic the effect of calculating higher orders in the new physics contributions too.\footnote{We have checked that, using a fixed scale $Q = m_t$, the $K$ factors so calculated are in good agreement with the ones obtained with a NLO calculation in Ref.~\cite{Franzosi:2015osa}.}
The new terms in the Lagrangian (\ref{ec:lagr}) are implemented in {\scshape Feynrules}~\cite{Alloul:2013bka} and interfaced to {\scshape MadGraph5} using the universal Feynrules output~\cite{Degrande:2011ua}. At each CM energy seven samples of $5 \times 10^5$ events are calculated for different values of $d_V$, to determine the quartic dependence of $\bar a_n$ with some redundancy and reduce the uncertainty from Monte Carlo statistics. For 8 TeV the fit yields, for the reference factorisation scale $Q$ equal to the total transverse mass,
\begin{align}
& \bar a_0~(\text{pb}) = 0.718 + 7.65 d_V +49.4 d_V^2 + 112 d_V^3 + 135 d_V^4 \,, \notag \\
& \bar a_1~(\text{pb}) = -0.158 -0.870 d_V -15.0 d_V^2 -49.9 d_V^3 -105 d_V^4 \,, \notag \\
& \bar a_2~(\text{pb}) = 0.0287 -0.120 d_V + 2.72 d_V^2 + 13.4 d_V^3 + 49.2 d_V^4 \,, \notag \\
& \bar a_3~(\text{pb}) = -0.0074 + 0.0724 d_V - 0.969 d_V^2 - 3.63 d_V^3 - 23.6 d_V^4 \,.
\label{ec:fit8dV}
\end{align}
normalised to the cross section for the $t \bar t \to \ell^+ \nu b \, \ell^- \bar \nu \bar b$ dilepton mode (no sum over leptons).
For the range of interest $|d_V| \lesssim 0.1$ the $d_V^3$ terms are subdominant and $d_V^4$ terms are numerically irrelevant. For 13 TeV we have
\begin{align}
& \bar a_0~(\text{pb}) = 2.39 +  25.5 d_V + 177 d_V^2 +  420 d_V^3 +  587 d_V^4 \,, \notag \\
& \bar a_1~(\text{pb}) = -0.573  -  3.36 d_V -  63.7 d_V^2 - 220 d_V^3 - 547 d_V^4 \,, \notag \\
& \bar a_2~(\text{pb}) = 0.114  - 0.360 d_V +  14.9 d_V^2 +  69.2 d_V^3 + 291 d_V^4 \,, \notag \\
& \bar a_3~(\text{fb}) = -0.0297 + 0.294 d_V -  4.46 d_V^2 - 22.2 d_V^3 -  166 d_V^4 \,.
\label{ec:fit13dV}
\end{align}
Again, $d_V^3$ terms are subdominant and $d_V^4$ terms can safely be neglected.
The scale uncertainty is estimated by setting the factorisation and renormalisation scales equal to twice and one half of the dynamic scale $Q$, and repeating the above procedure. The results for the normalised coefficients $a_1$, $a_2$ and $a_3$ are presented in Fig.~\ref{fig:avsdV}, for CM energies of 8 TeV and 13 TeV.
The uncertainty bands take into account the scale uncertainty and also the statistical Monte Carlo uncertainty. Furthermore, the bands are symmetrised around the reference predictions by taking the largest deviation with respect to the reference sample, in order to cover a possible bias in the fits (\ref{ec:fit8dV}) and (\ref{ec:fit13dV}) due to the Monte Carlo statistical uncertainty. 
 
Overall, we observe that the scale uncertainty in $a_1$, $a_2$ and $a_3$ is small. As anticipated, $a_1$ is the observable governing the effective spin correlation, which for $d_V$ small and positive increases up to $d_V \sim 0.04$, reaching an effective correlation $\fsm \sim 1.15$, and decreases for larger $d_V$. For negative $d_V$ the effective spin correlation decreases from the SM value. Since the 13 TeV ATLAS measurement $\fsm = 1.250 \pm 0.068$, is $3.7\sigma$ above the SM prediction, one expects tight constraints on negative values of $d_V$. Using this value of $\fsm$ as input, together with the dependence of the coefficients we have calculated, we estimate the limit $0.017 \leq d_V \leq 0.059$ at the 95\% confidence level (CL). As we have mentioned, with the 8 TeV dataset the CMS Collaboration already has obtained limits on $d_V$ from the normalised $d\sigma / d\phi$ distribution, $-0.027 \leq d_V \leq 0.021$ at the 95\% CL~\cite{Khachatryan:2016xws}. With 35.9 fb$^{-1}$ of 13 TeV data the limit using both the normalisation and the shape of the distribution is $-0.0018 \leq d_V \leq 0.012$~\cite{CMS:2018nzx}. For comparison, indirect limits from rare $B$ meson decays imply $-0.0012 \leq d_V \leq 0.0038$ at the 95\% CL~\cite{Martinez:2001qs}. The latter limits are quite model-dependent, however. A further study of this deviation and its potential explanation in terms of an anomalous top chromomagnetic moment is presented in section~\ref{sec:a}.

\section{Effect of a $tbW$ anomalous coupling}
\label{sec:4}

Although new physics in the top quark decay does not modify the $t \bar t$ spin correlation, it changes the spin analysing power of the the charged lepton $\alpha_\ell$ in (\ref{ec:tdec}), as well as the lepton energy distribution in the top quark rest frame, thereby modifying the dilepton azimuthal correlation. New physics in the $tbW$ vertex can affect several observables in the top quark decay, for example the $W$ polarisation fractions~\cite{Kane:1991bg,AguilarSaavedra:2010nx}, general $W$ spin observables~\cite{Aguilar-Saavedra:2015yza},  and the single top production cross sections~\cite{Chen:2005vr,Boos:1999dd,AguilarSaavedra:2008gt,Zhang:2010dr}, and are quite constrained by the measurement of those observables in top production and decay. However, there are `flat directions' in which the constraints are much looser. The $tbW$ interaction including contributions from dimension-six operators can be written as~\cite{AguilarSaavedra:2008zc}
\begin{eqnarray}
\mathcal{L}_{tbW} & = & - \frac{g}{\sqrt 2} \bar b \, \gamma^{\mu} \left( V_L
P_L + V_R P_R \right) t\; W_\mu^- \nonumber \\
& & - \frac{g}{\sqrt 2} \bar b \, \frac{i \sigma^{\mu \nu} q_\nu}{M_W}
\left( g_L P_L + g_R P_R \right) t\; W_\mu^- + \mathrm{h.c.} \,,
\label{ec:lagr2}
\end{eqnarray}
with $g$ the electroweak coupling, $M_W$ the $W$ boson mass, and $q = p_t - p_b$ its momentum; $V_L$ equals the Cabibbo-Kobayashi-Maskawa matrix element $V_{tb}$ in the SM, and $V_R$, $g_L$ and $g_R$ are anomalous couplings, which vanish in the SM at the tree level. One example of a flat direction is a combination of anomalous couplings with $M_W V_R = m_t g_L$. This combination can be generated by the redundant dimension-six operators~\cite{Buchmuller:1985jz}
\begin{align}
& O_{Dd}^{33} = (\bar q_{L3} D_\mu b_R) D^\mu \phi \,, 
& O_{\bar Dd}^{33} = (D_\mu \bar q_{L3} b_R) D^\mu \phi \,,
\end{align}
with $D_\mu$ the covariant derivative. The combination $O_{Dd}^{33} + O_{\bar Dd}^{33}$ does not contribute to the $t \to W^+ b$ amplitudes. The combination 
$O_{Dd}^{33} - O_{\bar Dd}^{33}$ generates an anomalous interaction~\cite{AguilarSaavedra:2008zc}
\begin{equation}
\mathcal{L}^\prime_{tbW} = - \frac{g}{\sqrt 2} h_L \bar b_R i \overleftrightarrow{\partial^\mu} t_L  W_\mu^+ \,.
\label{ec:hL}
\end{equation}
An interaction of this type only modifies the diagonal entry in the $W$ spin density matrix corresponding to $(0,0)$  helicities, therefore the constraints on $h_L$ are loose (see Ref.~\cite{Aguilar-Saavedra:2017wpl} for a detailed discussion). Moreover, a coupling $h_L$ in the $tbW$ vertex is equivalent to anomalous couplings $V_R = m_t h_L$, $g_L = M_W h_L$ in the minimal Lagrangian (\ref{ec:lagr2}), plus small terms proportional to the $b$ quark mass. Therefore, the insensitivity to the interaction (\ref{ec:hL}) is translated into a flat direction in the $(V_R,g_L)$ plane.

\begin{figure}[t]
\begin{center}
\begin{tabular}{ccc}
\includegraphics[height=4.8cm,clip=]{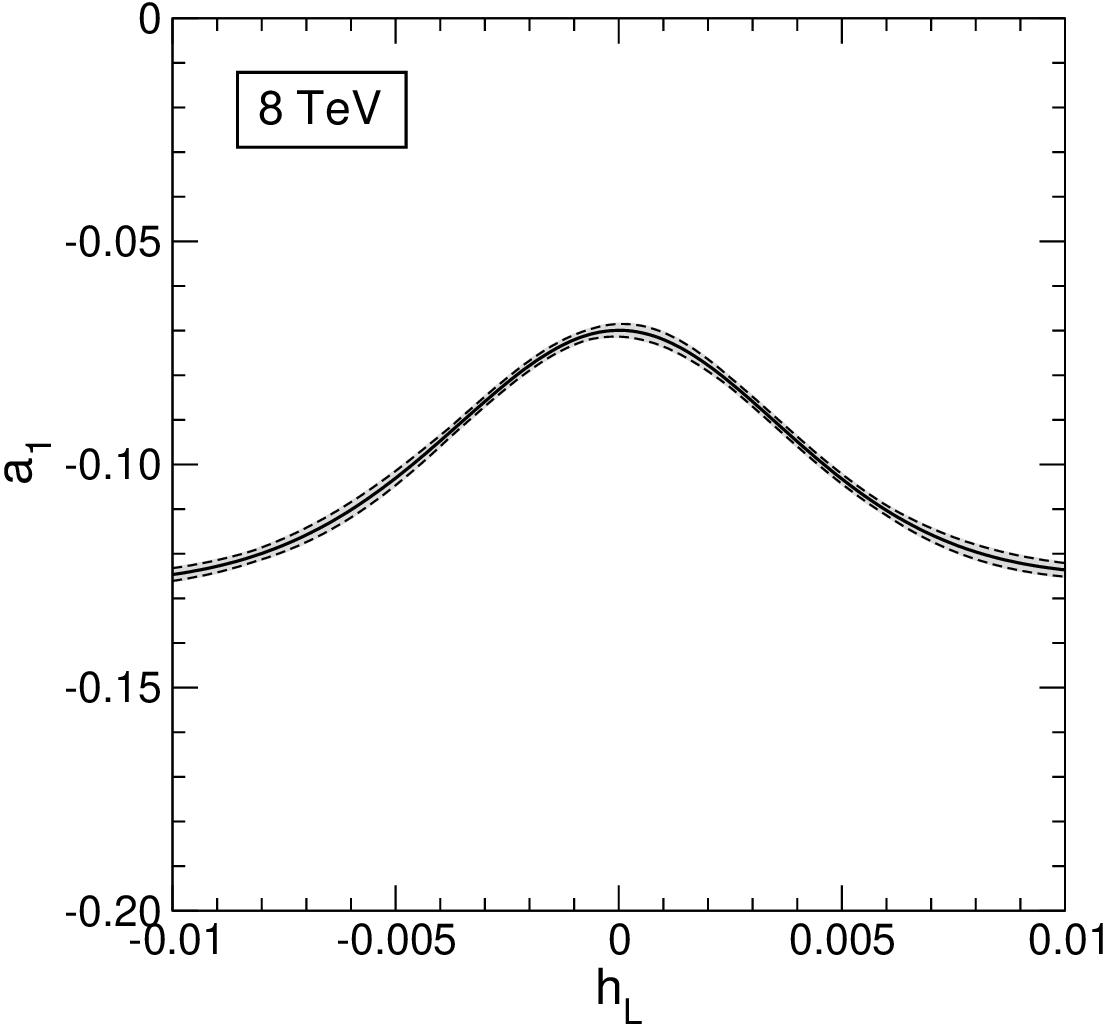}  &
\includegraphics[height=4.8cm,clip=]{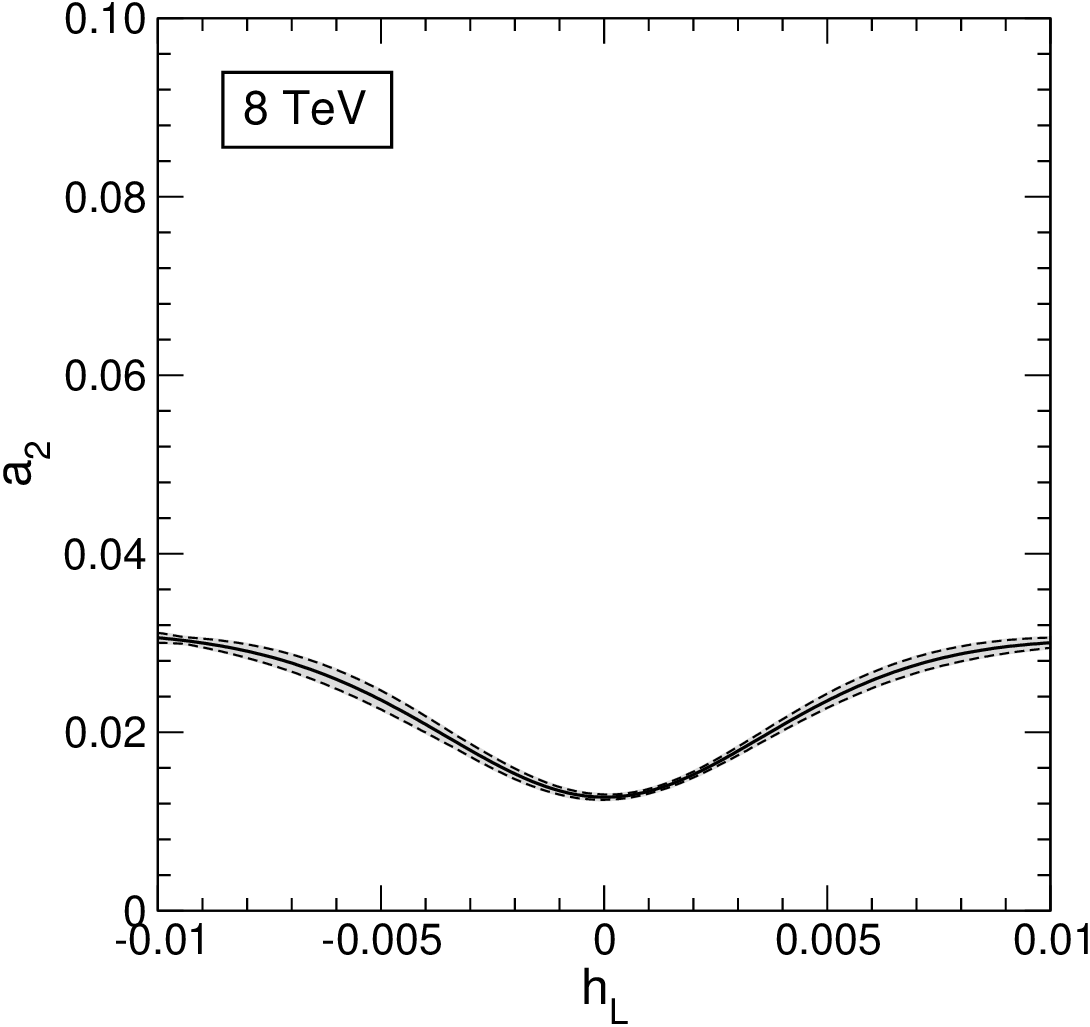}  &
\includegraphics[height=4.8cm,clip=]{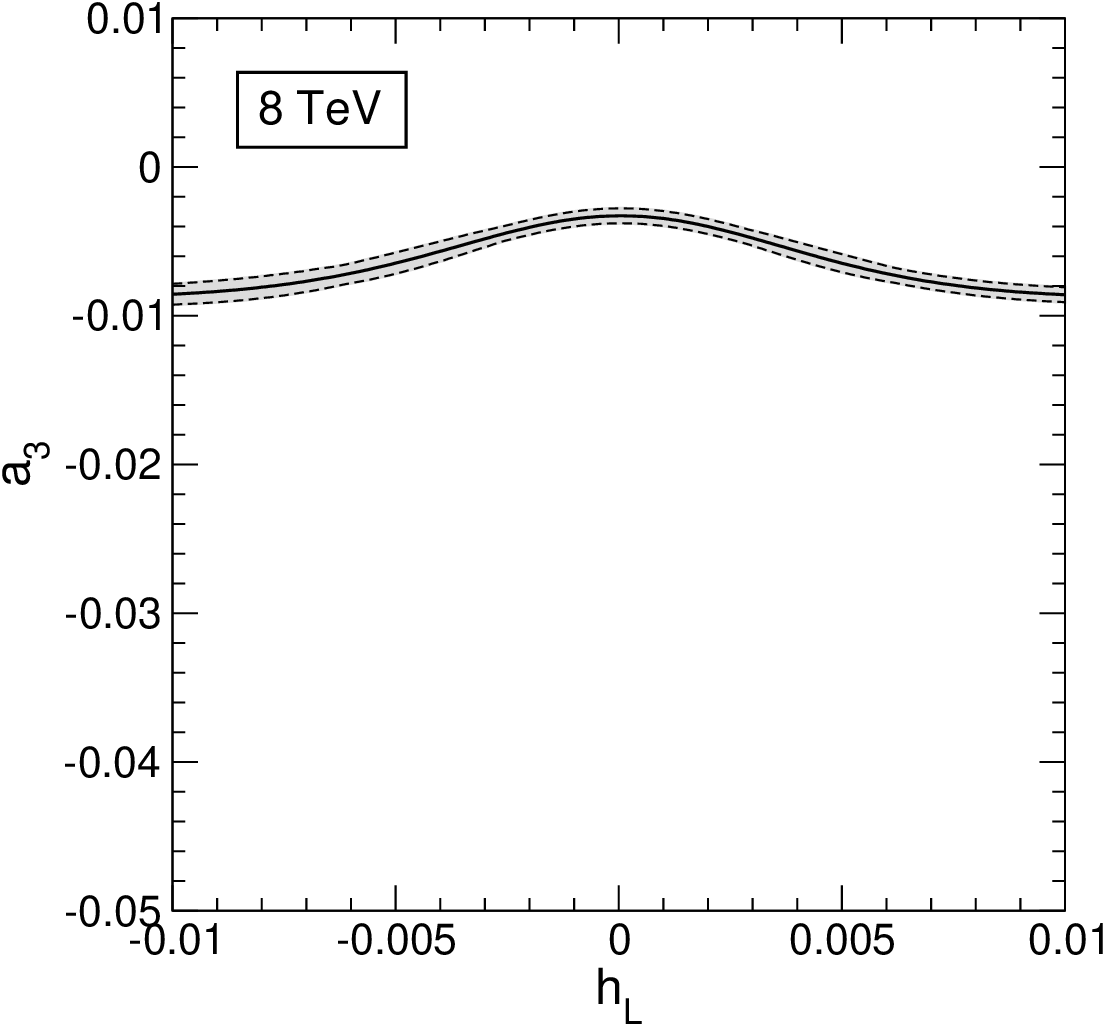}  \\
\includegraphics[height=4.8cm,clip=]{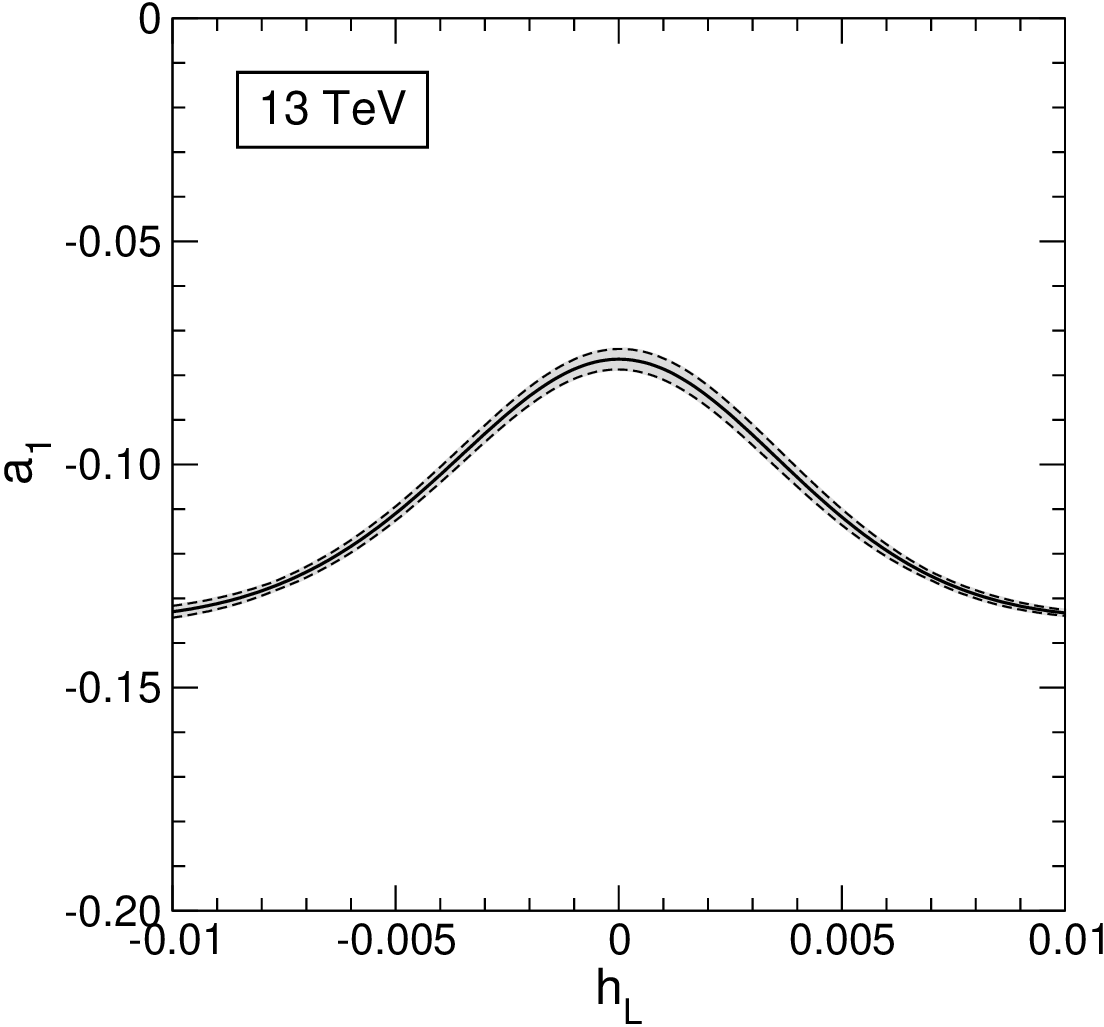}  &
\includegraphics[height=4.8cm,clip=]{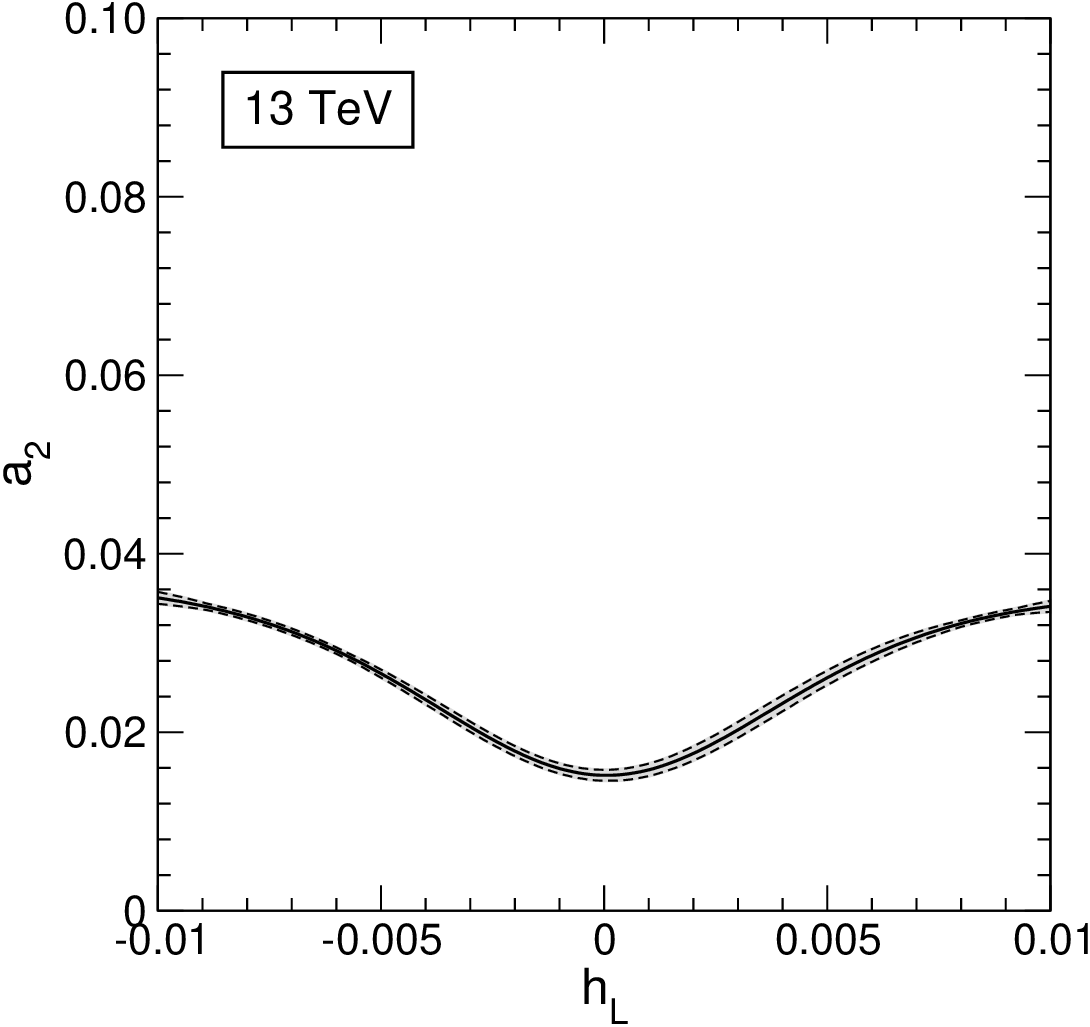}  &
\includegraphics[height=4.8cm,clip=]{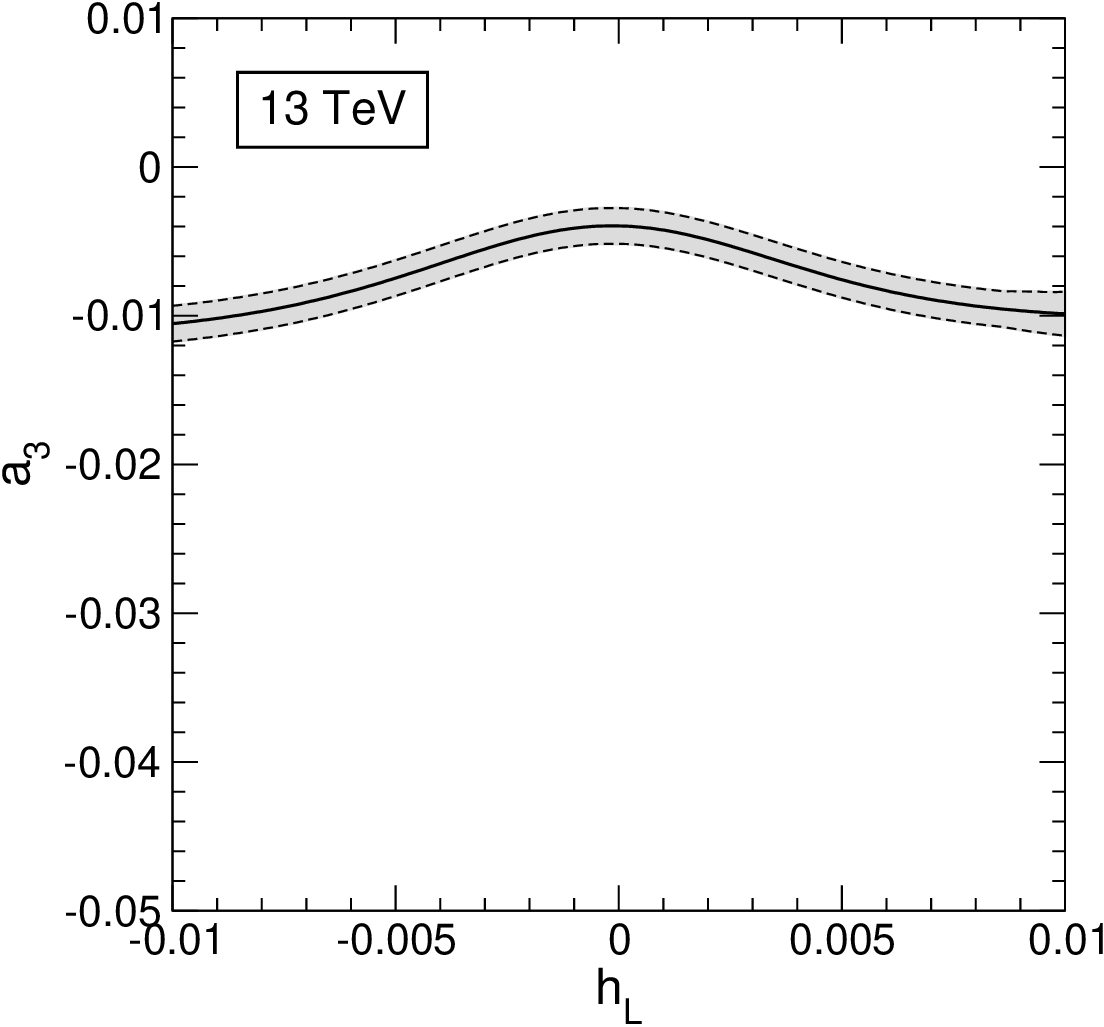}  
\end{tabular}
\caption{Dependence of the first Fourier coefficients in (\ref{ec:fou}) on a possible top $tbW$ anomalous coupling $h_L$ in Eq.~(\ref{ec:hL}), for CM energies of 8 TeV and 13 TeV.}
 \label{fig:avshL}
\end{center}
\end{figure}

We have followed the same procedure outlined in the previous section to calculate the dependence of $\bar a_n$ on the anomalous coupling $h_L$, with one minor difference. A non-zero $h_L$ changes the top width $\Gamma_t$, so that the production $\times$ decay cross section does not change in the narrow width approximation. Because we are interested in the normalised distribution, we can for simplicity keep $\Gamma_t$ fixed to its SM value in the calculations, in which case the dependence of the unnormalised $\bar a_n$ on $h_L$ is a fourth order polynomial. The difference with respect to the calculation with a varying $\Gamma_t$ is common for all $\bar a_n$, so it cancels when making the ratio to obtain the normalised quantities.
We generate seven samples for each CM energy, at the reference factorisation and renormalisation scale $Q$, and repeat the same for scales $Q \times 2$ and $Q / 2$ to estimate the scale uncertainty. At 8 TeV we find, defining the shorthand $\hat h_L = h_L/100$,
\begin{align}
& \bar a_0~(\text{pb}) = 0.718 - 0.0736 \hat h_L + 1.83 \hat h_L^2 - 0.0943 \hat h_L^3 + 1.17 \hat h_L^4 \,, \notag \\ 
& \bar a_1~(\text{pb}) = -0.158 +0.0171 \hat h_L - 0.887 \hat h_L^2 + 0.0545 \hat h_L^3 -0.408 \hat h_L^4 \,, \notag \\ \displaybreak 
& \bar a_2~(\text{pb}) = 0.0287  - 3.59 \times 10^{-3} \hat h_L + 0.232 \hat h_L^2 -0.0156 \hat h_L^3 + 0.0942 \hat h_L^4 \,, \notag \\
& \bar a_3~(\text{pb}) = -0.0074 + 1.18 \times 10^{-3} \hat h_L - 0.0650 \hat h_L^2 -3.14 \times 10^{-3} \hat h_L^3 - 0.0279 \hat h_L^4 \,.
\label{ec:fit8hL}
\end{align}
At 13 TeV we find
\begin{align}
& \bar a_0~(\text{pb}) = 2.39 - 0.243 \hat h_L + 6.08 \hat h_L^2 - 0.317 \hat h_L^3 + 3.90 \hat h_L^4 \,, \notag \\
& \bar a_1~(\text{pb}) = -0.572 +0.0601 \hat h_L - 3.16 \hat h_L^2 + 0.169 \hat h_L^3 -1.44 \hat h_L^4 \,, \notag \\
& \bar a_2~(\text{pb}) = 0.114  - 0.0172 \hat h_L + 0.815 \hat h_L^2 -0.0621 \hat h_L^3 + 0.416 \hat h_L^4 \,, \notag \\
& \bar a_3~(\text{pb}) = -0.0297 + 1.99 \times 10^{-3} \hat h_L - 0.242 \hat h_L^2 - 0.0327  \hat h_L^3 - 0.125 \hat h_L^4 \,.
\label{ec:fit13hL}
\end{align}
The quadratic and quartic terms are the dominant ones for the range of interest $|h_L| \lesssim 0.01$. We note that, despite the fact that at leading order $\alpha_\ell$ is not modified by new physics~\cite{Grzadkowski:1999iq,Rindani:2000jg,Grzadkowski:2002gt}, linear terms in $h_L$ appear in the above equations. These are justified by the potential dependence on $h_L$ of the lepton energy distribution in the top rest frame, which also affects the $d\sigma / d\phi$ distribution. The predictions for the normalised coefficients $a_1$, $a_2$ and $a_3$ are presented in Fig.~\ref{fig:avshL}.

The 13 TeV ATLAS measurement as well as previous ones indicate an enhanced effective correlation, therefore using the measurements of $\fsm$ as input to obtain constraints on $h_L$, which modifies the correlation in the opposite way, is not sensible. Instead, to estimate the potential to set limits on $h_L$ --- once the source of the present discrepancy is identified --- we can use $\fsm = 1 \pm 0.068$ as input, centred at the SM prediction and with the uncertainty of the ATLAS 13 TeV measurement. Proceeding this way we obtain a very tight limit, $|h_L| \leq 1.1 \times 10^{-3}$. 
 Translated into the couplings in the Lagrangian (\ref{ec:lagr2}), this amounts to $V_R \simeq 0.19$, $g_L \simeq 0.088$. The only existing limits covering this flat direction have been obtained by the ATLAS Collaboration in Ref.~\cite{Aaboud:2017yqf} with an analyis of triple differential angular distributions in top quark decays~\cite{Boudreau:2013yna}. The point $V_R = 0.19$, $g_L = 0.088$ is well within the $1\sigma$ allowed region, which extends up to $V_R \simeq 0.3$, $g_L \simeq 0.15$. This highlights the sensitivity of this distribution to probe anomalous contributions in the top decay, a possibilty that is yet unexplored.

\section{The 13 TeV anomaly on focus}
\label{sec:a}

\begin{figure}[t]
\begin{center}
\includegraphics[height=7cm,clip=]{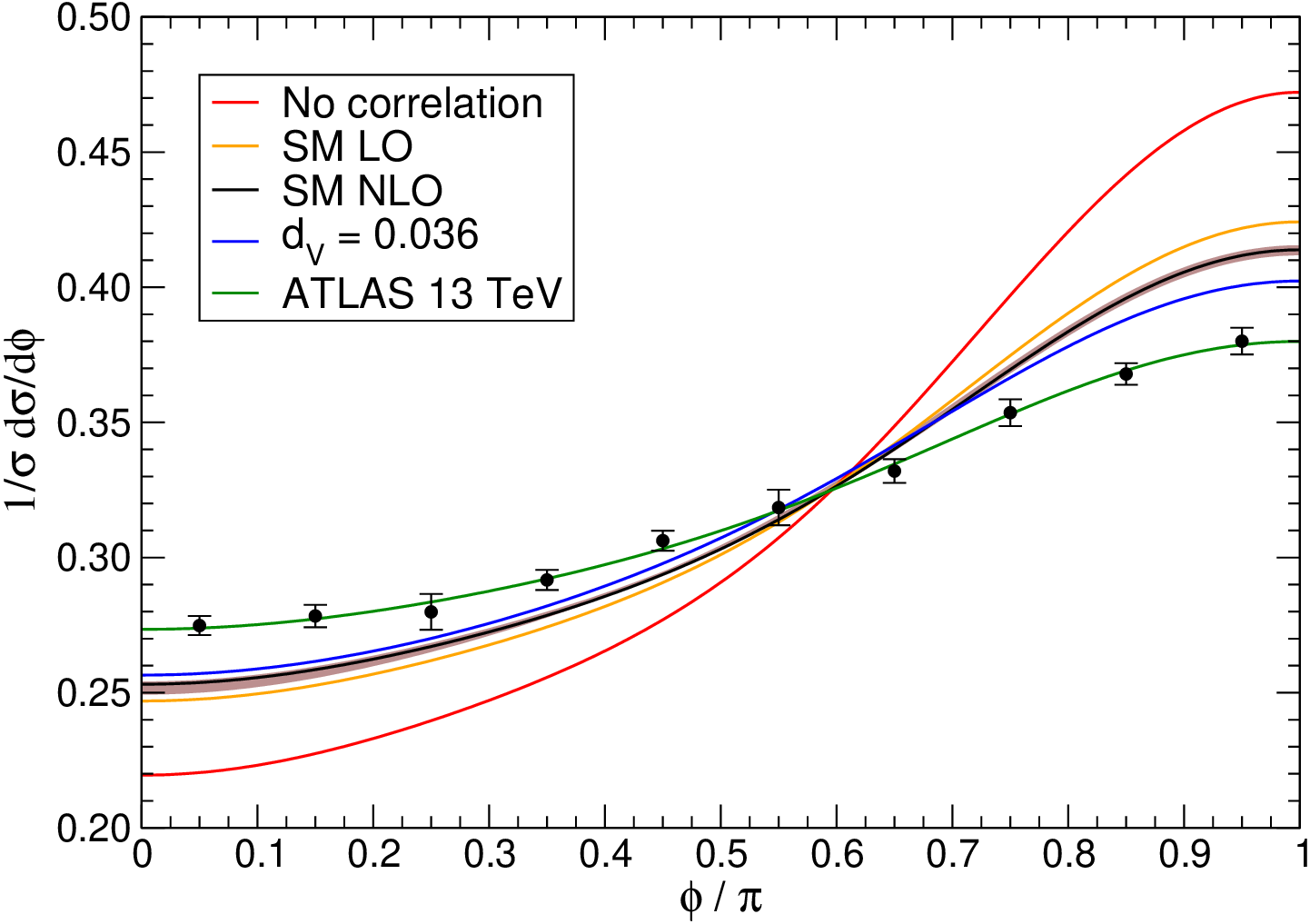} 
\caption{Different predictions for the azimuthal distribution at 13 TeV, compared to the ATLAS measurement.}
 \label{fig:phi13}
\end{center}
\end{figure}

The parameters characterising the $d\sigma/d\phi$ distribution measured by the ATLAS Collaboration can easily be obtained by digitising the plot. At the parton level, in full phase space, they are
\begin{equation}
a_1 = -0.0512 \,,\quad a_2 = 0.0082 \,,\quad a_3 = -0.0021 \,.
\end{equation}
and in fiducial phase space (with acceptance cuts)
\begin{equation}
a_1 = -0.123 \,,\quad a_2 = 0.0178 \,,\quad a_3 = -0.0021 \,.
\end{equation}
The determination of these coefficients with their correlation and uncertainties requires using the full event dataset, and has to be performed by the ATLAS Collaboration. Still, the values above obtained provide a good starting point to address possible explanations of the anomaly. Here we use the full phase space quantities to compare with put Monte Carlo predictions, noting that, although the corrections from the fiducial to full phase space are sizeable, the discrepancy is already observed in the former, c.f. Ref.~\cite{ATLAS:2018rgl}. The distributions presented in Fig.~\ref{fig:phi13} correspond to, from steeper to flatter slopes:
\begin{itemize}
\item[(a)] Uncorrelated $t \bar t$ production, calculated with {\scshape MCFM} using CT14 PDFs~\cite{Dulat:2015mca} and fixed factorisation and renormalisation scales equal to the top mass.
\item[(b)] The SM LO prediction, calculated with {\scshape Madgraph} using NN23 PDFs and factorisation and renormalisation scales the total transverse mass.
\item[(c)] The SM NLO prediction, also calculated with {\scshape Madgraph}, using the same scales. The shaded band around the line corresponds to the scale uncertainty obtained using one half and twice the total transverse mass. An alternative NLO prediction calculated with {\scshape MCFM} basically coincides with the central prediction calculated with {\scshape Madgraph} and is contained within the shaded band. It is not shown for clarity.
\item[(d)] The SM NLO prediction above plus a chromomagnetic coupling $d_V = 0.036$, which yields $a_1 = -0.0705$ (corresponding to $\fsm \simeq 1.15$). As we have mentioned, this is the maximum effective correlation that can be achieved in this context.
\item[(e)] The ATLAS result in Ref.~\cite{ATLAS:2018rgl}. The points correspond to the ATLAS data and their uncertainties.
\end{itemize}
The continuous lines in Fig.~\ref{fig:phi13} correspond to the third-order Fourier expansion in all cases.
As this plot highlights, the difference between data and the various predictions is quite significant. Some comments are in order.
\begin{enumerate}
\item The shift in the SM prediction from LO to NLO is much smaller than the difference between the NLO prediction and data, suggesting that the deviation is not due to higher-order corrections unaccounted for. The inclusion of $t \bar t j$ at the partonic level, matched with $t \bar t$, gives some enhancement of the effective correlation~\cite{ATLAS:2018rgl}, but the Monte Carlo predictions used by the ATLAS Collaboration still deviate from data by $3.2\sigma$.
\item The transverse momentum distributions of the top (anti-)quark measured by the CMS Collaboration also present discrepancies with respect to theory calculations~\cite{CMS:2018nzx} but these do not seem enough to explain the ATLAS deviation. By applying a crude linear reweighting factor of $1.07 - 4.8 \times 10^{-4}~\text{GeV}^{-1} \times p_T^t$, obtained from the top $p_T$ distribution in fiducial phase space in Ref.~\cite{CMS:2018nzx}, the NLO $d\sigma/d\phi$ distribution is mildly modified, from $a_1 = -0.0764$ to $a_1 = -0.0719$, still far from the ATLAS measurement.
\item An anomalous top chromomagnetic moment alone seems insufficient to explain the deviation observed. It remains to be seen how large the effective correlation can be by including $t \bar t$ plus jets in the presence of a non-zero $d_V$, higher-order corrections, etc. but reaching the observed distribution seems hard.
\end{enumerate}
With these caveats in mind, let us now discuss other possible effects and constraints on this potential explanation of the anomaly, also as a reference for other SM effects or new physics contributions that can modify the distribution.
An anomalous top chromomagnetic moment of the size $d_V \simeq 0.04$, so as to maximise the effective correlation, can be accommodated by the measurements of the total $t \bar t$ cross section, because the predictions are somewhat dependent on the factorisation and renormalisation scales. At 8 TeV our predictions are, for the reference dynamic scale $Q = \sum_i m_{Ti}$, twice, and one half of this value,
\begin{align}
&  \sigma~(\text{pb}) =  182 + 1950 d_V + 12600 d_V^2 && [Q] \,, \notag  \\ 
& \sigma~(\text{pb}) =  157 + 1670 d_V + 10600 d_V^2 && [Q \times 2] \,, \notag  \\ 
& \sigma~(\text{pb}) =  211 + 2250 d_V + 14700 d_V^2 && [Q / 2] \,,
\label{ec:xs8}
\end{align}
dropping terms of third and fourth order.
Using the combination of ATLAS and CMS $t \bar t$ cross section measurements in the $e\mu$ dilepton channel $\sigma_\text{exp} = 241.5 \pm 1.4\; (\text{stat}) \pm 5.7\; (\text{sys}) \pm 6.2\; (\text{lumi})$ pb~\cite{CMS:2014gta}, we find the loose constraint $0.006 \leq d_V \leq 0.046$ if we require the agreement of any of the predictions in Eqs.~(\ref{ec:xs8}) with this measurement, within two standard deviations. Next-to-next-to-leading corrections and soft-gluon resummation~\cite{Czakon:2013goa} increase the total cross section by around 8\%, relaxing the small tension between the SM predictions ($d_V = 0$) and the measurement, and strengthening the upper bound on $d_V$.  At 13 TeV our predictions for the cross section are
\begin{align}
&  \sigma~(\text{pb}) =  607 + 6480 d_V + 45100 d_V^2 && [Q] \,, \notag  \\ 
& \sigma~(\text{pb}) =  530 + 5650 d_V + 39000 d_V^2 && [Q \times 2] \,, \notag  \\ 
& \sigma~(\text{pb}) =  690 + 7360 d_V + 51500 d_V^2 && [Q / 2] \,.
\label{ec:xs13}
\end{align}
The naive average of the most precise 13 TeV measurements by the ATLAS~\cite{Aaboud:2016pbd} and CMS~\cite{Sirunyan:2017uhy} Collaborations is $\sigma_\text{exp} = 853 \pm 24$ pb. Requiring $2\sigma$ agreement of any of the predictions in Eqs.~(\ref{ec:xs13}) with this value, we obtain the constraint $0.014 \leq d_V \leq 0.048$. Again, the small tension with the SM predictions ($d_V = 0$) would be relaxed by including higher-order corrections, and the upper limit on $d_V$ would be tighter.

A non-zero $d_V$ also modifies the spin correlation coefficients $C$ in (\ref{ec:C}). We restrict ourselves to $C_{kk}$ in the helicity basis, for which the naive average of ATLAS and CMS measurements at 8 TeV (see Table~\ref{tab:C}) is 
$C_{kk} = 0.284 \pm 0.061$. Another distribution of interest is the polar angle $\theta_{ij}$ between the momenta of the decay products $i,j$, in the respective rest frame of the parent top (anti-)quark,
\begin{equation}
\frac{1}{\sigma} \frac{d\sigma}{d\!\cos \theta_{ij}}  = \frac{1}{2} \left( 1 - D \alpha_i  \alpha_j \cos \theta_{ij} \right) \,.
\label{ec:D}
\end{equation}
The spin correlation coefficient $D$ can be written in the basis of nine independent $C$ coefficients~\cite{Bernreuther:2015yna}, as
\begin{equation}
D = -\frac{1}{3} \left( C_{kk} + C_{rr} + C_{nn} \right) \,,
\label{ec:Drel}
\end{equation}
with $C_{rr}$ and $C_{nn}$ the diagonal spin correlation coefficients corresponding to axes orthogonal to the (anti-)top momentum $\vec k$, within the production plane ($\vec r$) and perpendicular to it ($\vec n$). The $D$ coefficient has been precisely measured by the CMS Collaboration at 8 TeV, yielding~\cite{Khachatryan:2016xws} $D = - 0.204 \pm 0.02\;(\text{stat}) \pm 0.024\;(\text{sys})$. The ATLAS Collaboration has not directly measured $D$ from the distribution (\ref{ec:D}), but it can be obtained from the measurement of the $C$ coefficients~\cite{Aaboud:2016bit} by using the relation (\ref{ec:Drel}). Ignoring the correlations between experimental uncertainties, we obtain $D = -0.229 \pm 0.060$. The naive average of these two measurements, $D = -0.209 \pm 0.028$, is dominated by the direct determination by the CMS Collaboration.

\begin{figure}[htb]
\begin{center}
\begin{tabular}{ccc}
\includegraphics[height=4.8cm,clip=]{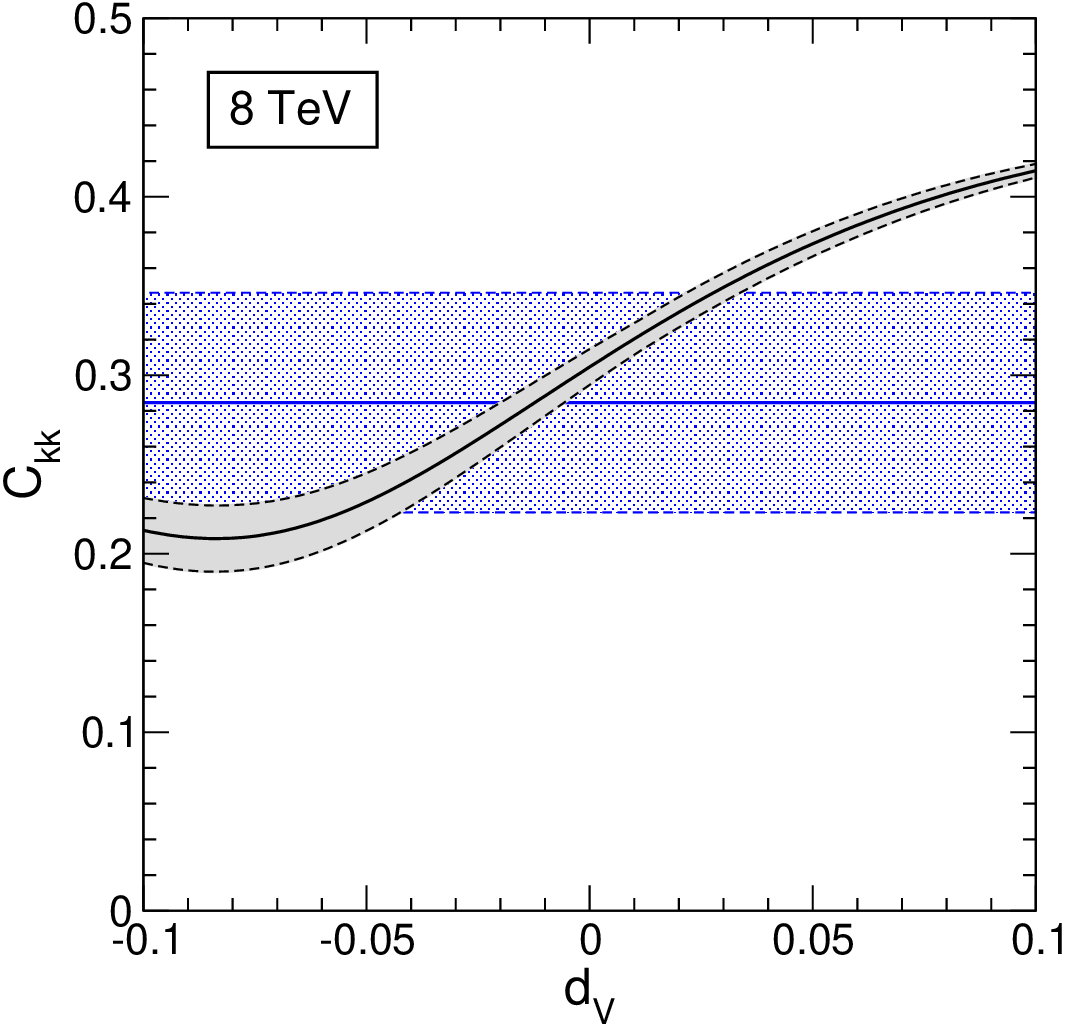}  &
\includegraphics[height=4.8cm,clip=]{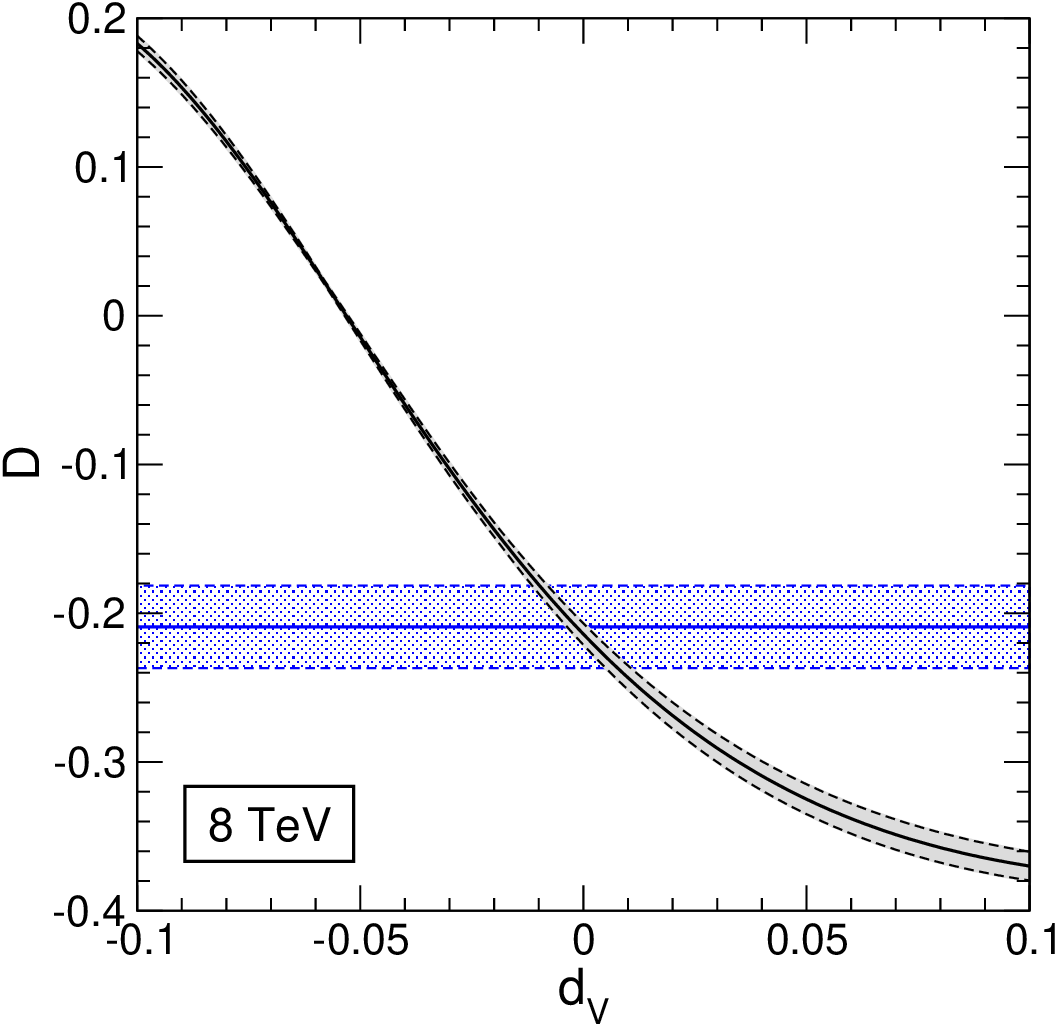}   \\
\includegraphics[height=4.8cm,clip=]{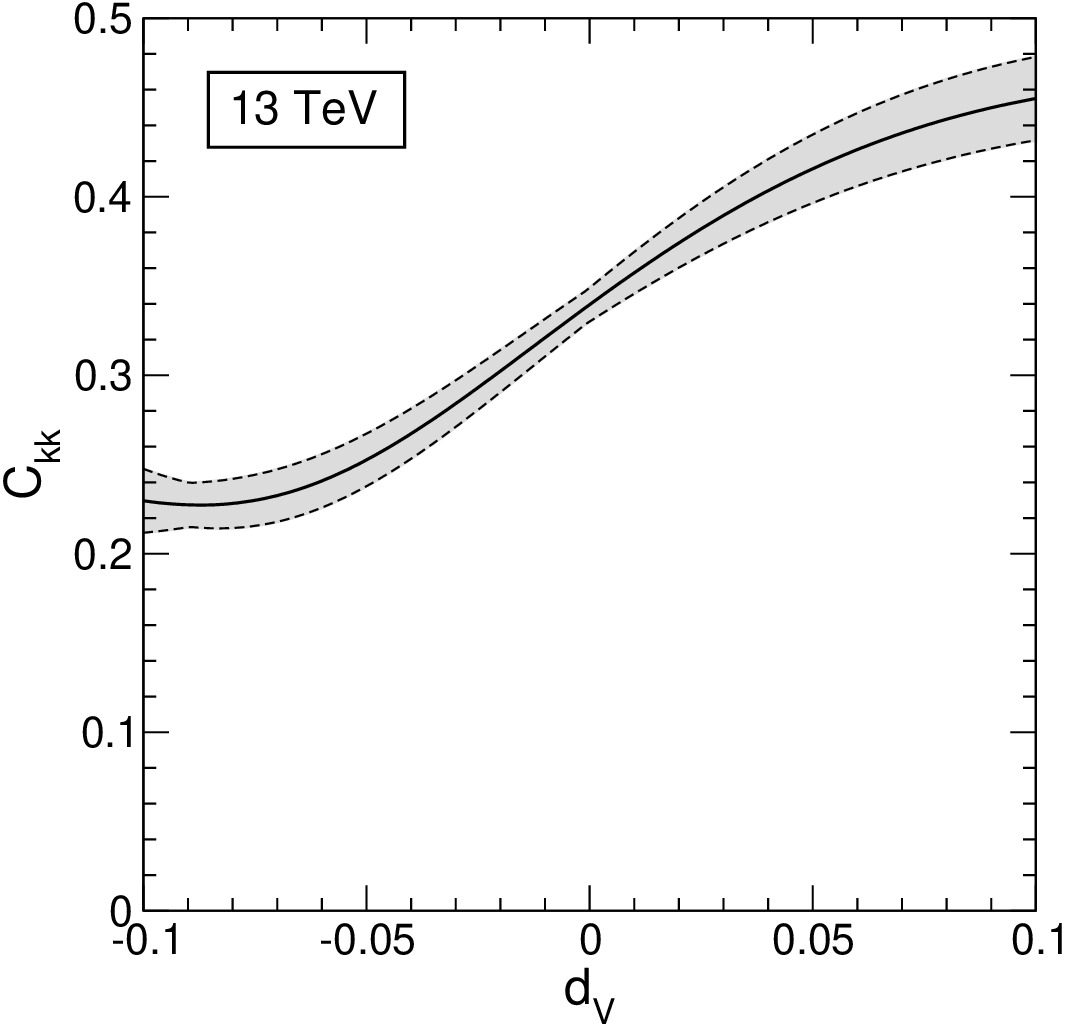}  &
\includegraphics[height=4.8cm,clip=]{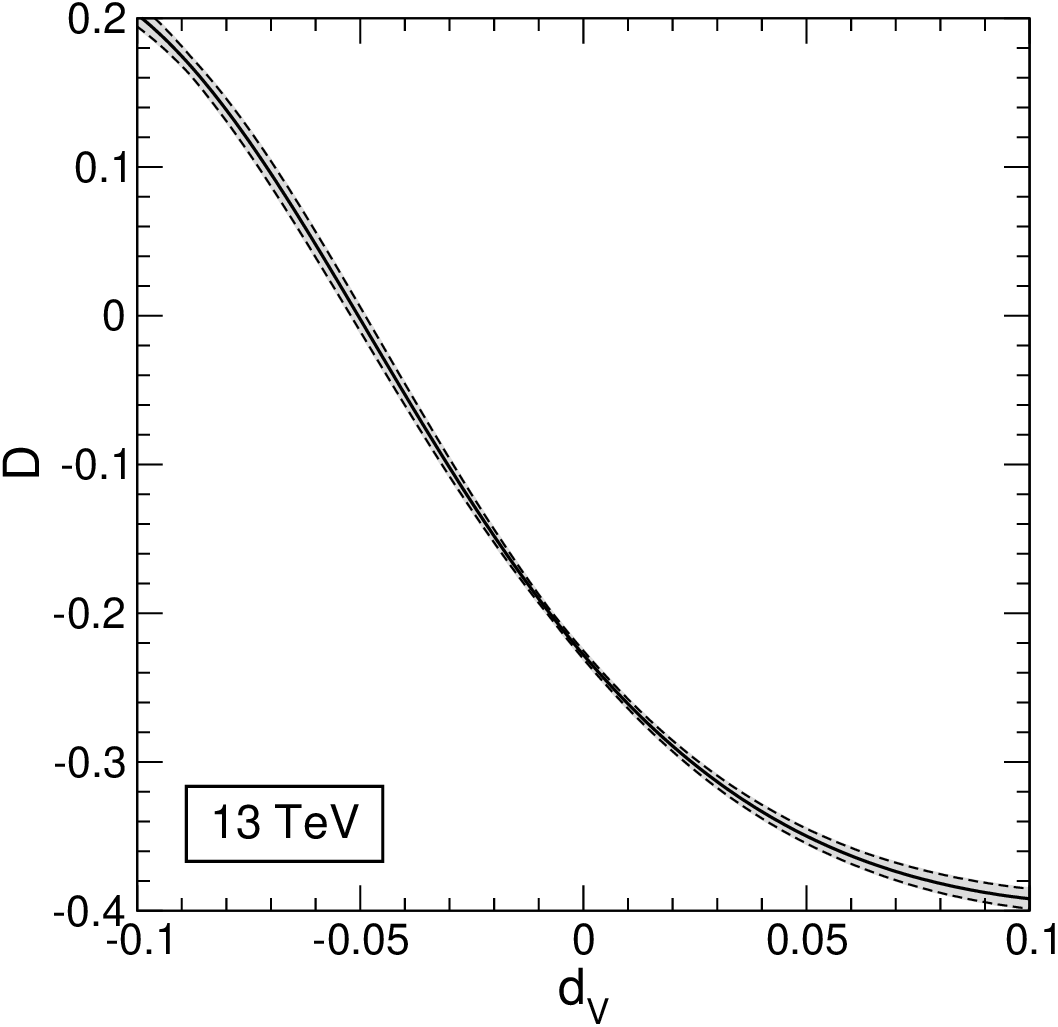} 
\end{tabular}
\caption{Dependence of the spin correlation coefficients $C_{kk}$ and $D$ on a possible top chromomagnetic moment $d_V$, for CM energies of 8 TeV and 13 TeV. The horizontal bands represent the naive averages of ATLAS and CMS measurements and their $1\sigma$ uncertainty.}
 \label{fig:CD}
\end{center}
\end{figure}

At 8 TeV our predictions for the reference scale $Q$ are
\begin{align}
& \sigma \times C_{kk}~(\text{pb}) = 55.6 + 886 d_V + 6500 d_V^2  \,, \notag \\
&  \sigma \times D~(\text{pb}) = -39.1 - 988 d_V - 5180 d_V^2  \,,
\end{align}
with $\sigma$ the total cross section in the first of Eqs.~(\ref{ec:xs8}). At 13 TeV we obtain
\begin{align}
& \sigma \times C_{kk}~(\text{pb}) = 206 + 3310 d_V + 25000 d_V^2  \,, \notag \\
&  \sigma \times D~(\text{pb}) = -138 - 3620 d_V - 18900 d_V^2 \,,
\end{align}
with $\sigma$ the total cross section in the first of Eqs.~(\ref{ec:xs13}). 
For the $d_V$ interval of interest, third and fourth order terms can safely be neglected at both CM energies. The dependence of $C_{kk}$ and $D$ on $d_V$ is depicted in Fig.~\ref{fig:CD}, with the uncertainty bands computed using scales $Q \times 2$ and $Q/2$, as in the previous sections. In the predictions for 8 TeV we include for comparison horizontal bands corresponding to the above obtained averages of experimental measurements, with their $1\sigma$ uncertainty. Because the relative contributions of a non-zero $d_V$ to $C_{rr}$ and $C_{nn}$ are larger~\cite{Bernreuther:2015yna} than the contribution to $C_{kk}$, the variation of $D$ is more pronounced. Therefore, we observe that, although the measurements of $C_{kk}$ are not very restrictive, the measurements of $D$ disfavour values of $d_V$ at the few percent level. Still, there is a caveat in the comparison because the measurement \cite{Khachatryan:2016xws} of $D$ includes the $e^+ e^-$ and $\mu^+ \mu^-$ dilepton channels, for which a cut on dilepton masses $m_{\ell \ell}$ around the $Z$ boson mass is applied to reduce the background from $Z$ plus jets. The presence of this cut at the reconstructed level, which affects events with smaller $\phi$, might bias the comparison of the unfolded measurement of $D$ with theory predictions in the presence of new physics. These arguments and caveats are expected to hold for other types of new physics yielding an enhanced spin correlation.
 
\section{Discussion}
\label{sec:5}

The use of a Fourier series to study the behaviour of a function is two centuries old. Still, series expansions have rarely been used in collider phenomenology to parameterise and scrutinise angular distributions. We advocate their use as a bridge between theory and experiments:
\begin{itemize}
\item[(i)] As a simple method to cast theory predictions for distributions that are otherwise difficult to parameterise. The series expansion allows the experiments to determine the full distribution with any desired binning.
\item[(ii)] As a simple output to report experimental measurements (including their correlation when necessary), allowing the theorists for easy reinterpretations by comparing the coefficients predicted by any model with the measured ones.
\end{itemize}
As an example of (i), we have considered the dilepton azimuthal correlation in $t \bar t$ production and its expansion as a Fourier series, finding that the number of coefficients required to determine the distribution is quite small. We have shown in sections~\ref{sec:3} and \ref{sec:4} that theory predictions for $d\sigma / d\phi$  including new physics in the production of $t \bar t$ pairs or in the top quark decay can easily be synthesised in terms of these coefficients. This allows for an easy study and comparison of different predictions with data, as done in section~\ref{sec:a}. It also allows the experiments to easily investigate potential new physics effects from the measurement of this angular distribution, by using the analytical dependence of the Fourier coefficients on the new physics couplings. The results in section~\ref{sec:4} are interesting on their own, as they show the good potential of this distribution to set limits on a flat direction in the parameter space of anomalous $tbW$ couplings.

A third advantage of a series expansion is the potential to pinpoint subtle deviations in the distributions, which might manifest in higher-order coefficients. These deviations might be caused not only by new physics, but also by detector effects. This possibility motivates the use of series expansions in the analysis of other angular distributions, even those that are easily parameterised and for which (i) and (ii) above are not needed. Let us consider for example the well-known angular distribution in top quark decays $t \to W^+ b \to \ell^+ \nu b$, corresponding to the angle $\theta_\ell^*$ between the momentum charged lepton momentum in the $W$ rest frame and the $W$ boson momentum in the top quark rest frame. The normalised distribution is
\begin{equation}
\frac{1}{\Gamma} \frac{d \Gamma}{d\!\cos \theta^*} = \frac{3}{8}
(1 + \cos \theta^*)^2 \, F_+ + \frac{3}{8} (1-\cos \theta^*)^2 \, F_-
+ \frac{3}{4} (1- \cos^2 \theta^*) \, F_0 \,,
\label{ec:dist2}
\end{equation}
with $F_+$, $F_-$ and $F_0$ the $W$ helicity fractions~\cite{Kane:1991bg}, satisfying $F_+ + F_- + F_0 = 1$. The functional form of this distribution is determined by angular momentum conservation, whereas the values of the helicity fractions are given by the interactions, for example $F_0 \simeq 0.703$, $F_- \simeq 0.297$, $F_+ \simeq 0$ in the SM at the tree level.
This distribution admits a finite expansion in Legendre polynomials as written in (\ref{ec:leg}) but with $x \equiv \cos \theta^*$. The non-zero coefficients are
\begin{align}
& b_0 = \frac{1}{2} \,, \quad
b_1 = \frac{3}{4} \left( F_+ - F_- \right) \,, \quad
 b_2 = \frac{1}{4} \left( 1 - 3 F_0 \right) \,.
\end{align}
However, higher-order coefficients can be generated by detector effects. For example, we have verified that a deficient reconstruction of the $W$ rest frame, arising from a mismeasurement of the missing energy from the neutrino, can generate non-zero $b_4$, $b_5$, etc. With the high statistics that will be available at the LHC Run 2, it is of the utmost importance to have under very good control the signal modeling, data unfolding, etc. in order to perform precision physics. A series expansion, as proposed in this work, can reveal subtle effects and may become a very useful tool in order to test the robustness of the modeling, especially in case any deviation from the SM is found, as it may be the case with the 13 TeV ATLAS measurement of the azimuthal correlation.

Last, but not least, we have used the proposed framework to investigate in detail this anomaly, within the SM and in the presence of an enhanced top chromomagnetic coupling. We find that the deviation of data from SM predictions is unlikely to be due to missing higher-order corrections, and to explain this deviation in terms of an anomalous coupling is also difficult, though further work in this direction is required.

\section*{Acknowledgements}
This work has been supported by MINECO Project  FPA 2013-47836-C3-2-P (including ERDF).

\end{document}